\documentclass{aa}
\usepackage{graphicx,times,amssymb} \usepackage[english]{babel}
\newcommand{\etal}{et~al.}
\newcommand{\e}{et~al.}

\newcommand{\civ}{\ion{C}{IV}}
\newcommand{\cii}{\ion{C}{II}}
\newcommand{\ci}{\ion{C}{I}}
\newcommand{\oi}{\ion{O}{I}}
\newcommand{\mgii}{\ion{Mg}{II}}
\newcommand{\feii}{\ion{Fe}{II}}
\newcommand{\fei}{\ion{Fe}{I}}
\newcommand{\mgi}{\ion{Mg}{I}}
\newcommand{\siii}{\ion{Si}{II}}
\newcommand{\sii}{\ion{Si}{I}}
\newcommand{\siiv}{\ion{Si}{IV}}
\newcommand{\alii}{\ion{Al}{II}}

\def\be{\begin{equation}} \def\ee{\end{equation}}
\def\bea{\begin{eqnarray}} \def\eea{\end{eqnarray}}
  
  \def\e{et~al.}

\begin{document}

\title{\civ\ Absorbers in $z>4$ Quasars:\\
 Tracing Early Galactic Halos Evolution}

\author{ C\'eline P\'eroux\inst{1}\thanks{Marie Curie Fellow. E-mail:
peroux@ts.astro.it}, Patrick Petitjean\inst{2,3}, Bastien
Aracil\inst{2}, Mike Irwin\inst{4} \& Richard G. McMahon\inst{4} }

\institute{
$^1$Osservatorio Astronomico di Trieste, Via Tiepolo, 11, 34 131
Trieste, Italy\\
$^2$Institut d'Astrophysique de Paris, 98 bis Bld Arago, 75 014 
Paris, France\\
$^3$LERMA, Observatoire de Paris, 61 avenue de l'Observatoire, 
75 014 Paris, France\\
$^4$Institute of Astronomy, Madingley Road, Cambridge CB3 0HA, 
UK}

\date{\today} \authorrunning{P\'eroux}
\titlerunning{\civ\ Absorbers in $z>4$ Quasars}

\abstract {We use 29 $z>4$ quasar spectra to build a homogeneous
sample of high-redshift \civ\ absorbers. We use these data to
calculate the number density, $n(z)$, of $W_{\rm rest}(\civ)>0.30$\AA\
\civ\ doublets. We find that $n(z)$ increases with time from
$z\sim4.5$. In addition, $W_{\rm rest}(\civ)>0.15$\AA\ \civ\ systems
are more numerous than the former at all redshifts. On the contrary,
$n(z)$ of $W_{\rm rest}(\mgii)>0.30$\AA\ \mgii\ doublets decreases
with time, in agreement with results from studies of LLS number
densities. Below $z<3$, none of all the classes of absorbers show
signs of evolution. We interpret this as the formation of galactic
envelopes from smaller halos. Furthermore, the doublet ratio
DR=$W(\civ1548)$/$W(\civ1550)$ is found to decrease with time, a
signature of the increase of the mean \civ\ column density. Finally,
the $W$(\siiv1393)/$W$(\civ1548) ratios in strong absorbers ($W_{\rm
rest}(\civ1548) \ge 0.50$ \AA) is found to be approximately constant
from $z=3.5$ to $z=2.5$ and then to decrease with time. This result
suggests that the highest column density absorbers are not sensitive
to changes in the ionization continuum.

\keywords{Cosmology: observations -- Galaxies: evolution
-- Quasars: absorption lines -- Galaxies: halos -- Galaxies: high-redshift}}

\maketitle
\section{Introduction}

Various lines of evidence point towards a direct connection between
metal-line absorption systems observed in the spectra of high redshift
background quasars and normal galaxies. \mgii\ systems are known to be
associated with the gaseous envelopes of bright galaxies which have
been detected in emission at $z \sim 0.6$ (Bergeron \& Boiss\'e
1991). The typical impact parameter from the galaxy varies with the
equivalent width threshold of the surveys for absorbers: 40 h$^{-1}$
kpc at $W_{\rm rest}=0.30$\AA\ and 60 h$^{-1}$ kpc at $W_{\rm
rest}=0.02$\AA\ (Churchill et al. 1999). In addition the absorbers
optically thick to hydrogen ionizing radiation, i.e. neutral hydrogen
absorbers with N(HI)$>10^{17.2}$ cm$^{-2}$, the Lyman limit systems
(hereafter LLS) have sizes comparable to the $W_{\rm rest}=0.30$\AA\
\mgii\ selected systems. Therefore both \mgii\ and LLS are believed to
arise from the same type of object.

At $z<1$, \civ\ absorbers arise from the extended (100 h$^{-1}$ kpc)
halos of galaxies of a wide range of luminosity and morphological type
(Chen, Lanzetta \& Webb 2001).  More specifically, Adelberger et
al. (2003) have shown that at $z\sim 3$, all \civ\ absorbers are
situated within 200 h$^{-1}$ kpc of a Lyman-break galaxy and argue
that this distance is characteristic of a galactic-scale outflow
driven by star formation activity in this type of galaxy. Moreover,
simulations have shown that the observed \civ\ kinematic structure and
column densities can be well reproduced by merging of proto-galactic
clumps (Haehnelt, Steinmetz \& Rauch 1996). Compact halos of hot gas
with temperature close to $\sim 10^5$~K satisfactorily explain the
observed multi-component nature of the \civ\ absorbers. These
absorption systems thus provide a powerful observational tool to
understand the processes of galaxy formation and evolution that
deposit chemically enriched material far from the central galaxy.

The methods used to study the properties of \civ\ systems can be split
into two different approaches. First, observing \civ\ in high
signal-to-noise ratio, high-resolution spectra, allows us to determine
their column density down to log N(\civ) $\sim 11.8$ cm$^{-2}$. This
enables detailed kinematic and temperature studies (Rauch et al. 1996)
which show that \civ\ components may be the building blocks of future
normal galaxies. Such data are also used to determine the low end of
the \civ\ column density distribution (Ellison et al. 2000) and mass
density (Songaila 2001; Pettini et al. 2003; Boksenberg, Sargent \&
Rauch 2003) and to study the velocity structure within the halos
(e.g. Petitjean \& Bergeron 1994 and Crotts, Burles \& Tytler
1997). Nevertheless, such analysis are currently limited to a few
lines of sight.

The second approach consists in studying a statistically significant
number of absorbers by constructing a large homogeneous
sample. Steidel (1990) uses quasar spectra at $3.04<z_{\rm em}<4.11$
in addition to data at $1.08<z_{\rm em}<3.56$ from Sargent, Boksenberg
\& Steidel (1988) to build such a homogeneous sample. They find that
the number density of \civ\ systems, $n(z)$, increases with increasing
redshift in the range $z_{\rm abs}=1,3-3.7$ and deduce from this that
the properties of the absorbers are evolving with time. Using a
two-point correlation function to study the clustering of the \civ\
doublets, they find marginal evidence that the width of the
correlation peak on small velocity scales is smaller at high redshift
than in the local Universe. Quashnock, Vanden Berk \& York (1996) show
that clustering is also present at higher redshift, although its scale
does not appear to have changed significantly since then. More
recently, Misawa et al. (2002) have used higher-resolution spectra (2
\AA\ FWHM) from the Keck telescope to conduct a systematic search for
\civ\ doublets. They measure $n(z)$ up $z\sim4$ and find that their
results agree with Steidel (1990) at a rest equivalent width limit of
$W_{\rm rest}=0.30$~\AA, but not at $W_{\rm rest}=0.15$~\AA.  We
propose here that this discrepancy is actually due to incompleteness
in the Steidel sample at $W_{\rm rest}>0.15$~\AA\ linked with the
lower resolution of the survey and pay particular attention to this
point in our analysis.

Although both these approaches are complementary, the paucity of
strong \civ\ systems means that the second technique is more
appropriate to study the characteristics and evolution of the galactic
halos. In the present study, we use 29 $z>4$ quasar spectra to build a
new and homogeneous sample of high-redshift \civ\ absorbers. In
section 2, we summarise the set-up for the observations and detail the
data reduction process in section 3. The method for the identification
of the absorbers is presented in section 4 together with a list of
\civ\ systems. The statistical properties of this sample are derived
in section 5 and discussed in section 6. This paper assumes
$\Omega_{\Lambda}=0.7$ and $\Omega_0=0.3$ throughout.

\begin{table*}[t]
\begin{center}
\begin{tabular}{lcccccc}
\hline 
\hline 
Quasar &$z_{\rm em}$ &APM &Telescope &Date &Exp. Time &Spectral\\
Name   &             &R Mag&         &     &(sec)     &Coverage\\  
\hline 
BR  J0006$-$6208   &4.455  &18.3  &NTT  &2002 Aug	&9000	&7008$-$8512 \\
BR  J0030$-$5129   &4.174  &18.6  &NTT  &2002 Aug	&7500	&6508$-$8016 \\
PSS J0034$+$1639   &4.293  &18.0  &NTT  &2002 Aug	&6000	&6611$-$8117 \\
PSS J0133$+$0400   &4.154  &18.3  &NTT  &2002 Aug	&9000	&6365$-$7860 \\
PSS J0248$+$1802   &4.422  &17.7  &WHT  &2001 Nov	&3300	&6631$-$7442 \\
BR  J0311$-$1722   &4.039  &17.7  &NTT  &2001 Dec	&3600	&6500$-$7820 \\
BR  J0324$-$2918   &4.622  &18.7  &NTT  &2001 Dec 	&10800	&7150$-$8470 \\
BR  J0334$-$1612   &4.363  &17.9  &WHT  &2001 Nov	&1800	&6631$-$7442 \\
BR  J0355$-$3811   &4.545  &17.9  &NTT  &2001 Dec	&3600	&7150$-$8470 \\
BR  J0419$-$5716   &4.461  &17.8  &NTT  &2001 Dec	&5400	&7150$-$8470 \\
BR  J0426$-$2202   &4.320  &17.9  &NTT  &2001 Dec	&6000	&6930$-$8250 \\
PMN J0525$-$3343   &4.383  &18.4  &NTT  &2001 Dec	&9000	&6930$-$8250 \\
BR  J0529$-$3552   &4.172  &18.3  &NTT  &2001 Dec	&9000	&6500$-$7820 \\
BR  J0714$-$6455   &4.462  &18.3  &NTT  &2001 Dec	&9000	&7150$-$8470 \\
PSS J0747$+$4434   &4.430  &18.4  &WHT  &2001 Nov	&1200 	&6631$-$7442 \\
PSS J1159$+$1337   &4.073  &17.1  &WHT  &2002 Jun	&4500	&6000$-$7400 \\
PSS J1253$-$0228   &4.007  &18.8  &WHT  &2002 Jun	&6300	&6000$-$7400 \\
PSS J1330$-$2522   &3.949  &18.5  &WHT  &2002 Jun	&5400 	&6000$-$7400 \\
BR  J1456$+$2007   &4.249  &18.2  &WHT  &2002 Jun	&7200 	&6850$-$8350 \\
BR  J1618$+$4125   &4.213  &18.5  &WHT  &2002 Jun	&7200 	&6850$-$8350 \\
PSS J1633$+$1411   &4.351  &18.7  &NTT  &2002 Aug	&10800	&7008$-$8512 \\
PSS J1646$+$5514   &4.037  &17.1  &WHT  &2002 Jun	&4500   &6850$-$8350 \\
PSS J1723$+$2243$^*$&4.520 &18.2  &NTT  &2002 Aug	&3600  	&7008$-$8512 \\
PSS J1802$+$5616   &4.158  &18.3  &WHT  &2002 Jun	&6300  	&6850$-$8350 \\
PSS J2154$+$0335   &4.363  &19.0  &NTT  &2002 Aug	&10800	&6615$-$8117 \\
PSS J2155$+$1358   &4.256  &18.0  &WHT  &2002 Jun	&6300	&6850$-$8350 \\
BR  J2216$-$6714   &4.469  &18.6  &NTT  &2002 Aug	&10200	&7008$-$8512 \\
PSS J2344$+$0342   &4.239  &18.2  &NTT  &2002 Aug	&9600	&6510$-$8010 \\
BR  J2349$-$3712   &4.208  &18.7  &NTT  &2002 Aug	&10200	&6510$-$8010 \\
\hline\hline
\end{tabular}
\end{center}
\scriptsize{$^*$Object possibly affected by BAL features.}
\caption{Journal of Observations.}
\label{t:JoO}
\end{table*}

\section{Observations}

All the quasars observed in the present study were selected from the
lower-resolution survey of P\'eroux \e\ (2001). The new observations
were carried out during two observing runs at the 4.2 m William
Herschel telescope (WHT) of the Isaac Newton Group of telescopes in
the Canary Islands and two observing runs at the 3.58 m New Technology
Telescope (NTT) of the European Southern Observatory in La Silla,
Chile.  High signal-to-noise optical spectrophotometry was obtained
covering approximately 7000 \AA\space to 8000 \AA, the exact range
depending on which instrument was used for the observations.  A
journal of the observations is presented in Table~\ref{t:JoO}.

Eleven quasars were observed at the WHT during 2001 November 7 -- 8
and 2002 June 1 -- 2. The first run suffered bad weather conditions,
thus limiting the number of objects observed and the quality of the
resulting spectra. The integration times ranged from 1200 -- 7200
seconds. We used only the red arm of the ISIS spectrograph. A new CCD
was installed before the second run took place and therefore the
characteristics of the set-ups slightly changed between the two
observational sessions. For the November run a thinned coated
Tektronix CCD was used, while a thinned coated EEV CCD was used
for the June run. Grating R600R was used leading to a dispersion of
0.79 \AA\space pixel$^{-1}$ for the November run and 0.44 \AA\space
pixel$^{-1}$ for the June run. In all cases, a GG495 blocking filter
was used to minimize the second order contamination. All the
observations were taken with a slit width of 1.0 -- 1.5
arcsec. Blind-offsetting from bright $\sim 15 $--$ 17^{th}$ magnitude
stellar fiducials was used to position the quasars in the slit, partly
to save acquisition time and partly because the majority of the
quasars were not visible using the blue sensitive TV acquisition
system. Readout time was reduced by windowing the CCDs in the spatial
direction.

Eighteen quasars (including the possibly BAL quasar PSS J1723$+$2243)
were observed at the NTT during 2001 December 16 -- 18 and 2002 August
11 -- 13 using the ESO Multi-Mode Instrument (EMMI).  The exposure
times ranged from 3600 to 10800 seconds.  Again, a new CCD was
installed before the second run took place and therefore the
characteristics of the set-ups slightly changed between the two
observational sessions. For the December run a thin, back-illuminated
Tektronix CCD was used, while a MIT/LL CCD was used for the August
run. Grating \#7 used on both occassions and gave dispersions of 0.66
\AA\space pixel$^{-1}$ for the December run and 0.82 \AA\space
pixel$^{-1}$ for the August run. In both cases a OG530\#645 blocking
filter was used to minimize the second order contamination. All the
observations were taken with a slit width of $\sim$ 1.0
arcsec. Readout time was again reduced by windowing the CCDs in the
spatial direction.

\section{Data Reduction}

The data reduction was undertaken using the IRAF~\footnote{IRAF is
distributed by the National Optical Astronomy Observatories, which are
operated by the Association of Universities for Research in Astronomy,
Inc., under cooperative agreement with the National Science
Foundation.} software package. Because the bias frames for each nights
were so similar, a master `zero' frame for each run was created using
the IMCOMBINE routine.  The data were overscan-corrected,
zero-corrected, and trimmed using CCDPROC.  Similarly, a single
flat-field frame was produced by taking the median of the (suitably
scaled) flats. The overall background variation across this frame was
removed to produce an image to correct for the pixel-to-pixel
sensitivity variation of the data.  The task APALL was used to extract
1-D multi-spectra from the 2-D frames. The routine estimates the sky
level by model fitting over specified regions on either side of the
spectrum. In the cases where the objects were faint on the 2-D frames,
different exposures were IMCOMBINEd before extraction with APALL.

The WHT data were wavelength calibrated using CuAr and CuNe arcs and
the NTT data using HeAr and Ar lamps.  Night sky lines were used to
monitor the wavelength calibration for all spectra. In all but one
run, observations of B-stars free of strong features in the red were
taken in order to remove the effects of atmospheric absorption
(e.g. O$_2$ A band at 7600 \AA) in the quasar spectra. No such
observations were made in the December run at NTT, and therefore in
this case appropriately scaled B-star spectra were used for
atmospheric correction. The atmospheric absorption features seen in
the B-star spectrum were isolated by interpolating between values on
either side of the feature. The original B-star spectrum was then
divided by this atmospheric-free spectrum to create an atmospheric
correction spectrum. Finally the object spectra were divided by the
scaled correction spectrum.

The resolution of the resulting spectra ranges from 1.60 to
2.30\AA. An estimate of the continuum signal-to-noise ratio is given
in Table~\ref{t:sn} for each object. The spectra are shown in
Figure~\ref{f:spec} with arbitrary flux scales. The quasar redshifts
and magnitudes listed in Table~\ref{t:JoO} are from the original
survey of P\'eroux \e\ (2001). All the spectra were then continuum
normalised within the MIDAS data reduction software package (Fontana
\& Ballester 1995) in order to detect the signature of intervening
absorption lines. For this purpose a spline function was fitted to the
quasar continuum to smoothly connect the regions free from absorption
features, as is standard practice.

\begin{figure*}[t]
\includegraphics[width=.97\hsize,height=2.01\vsize]{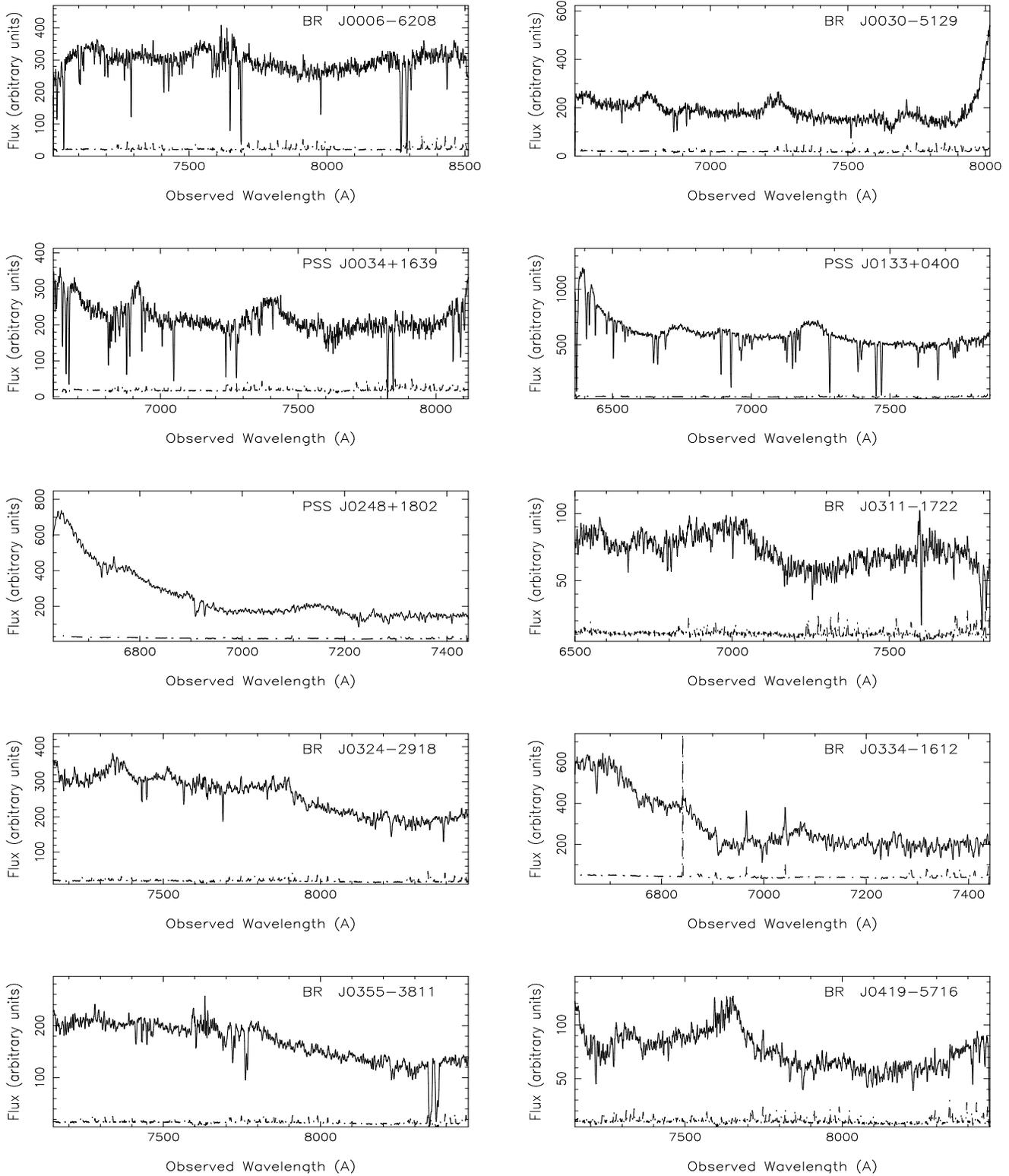}
\caption{Spectra of the 29 $z>4$ quasars observed as part of the survey.
\label{f:spec}}
\end{figure*}

\begin{figure*}[t]
\includegraphics[width=.97\hsize,height=2.01\vsize]{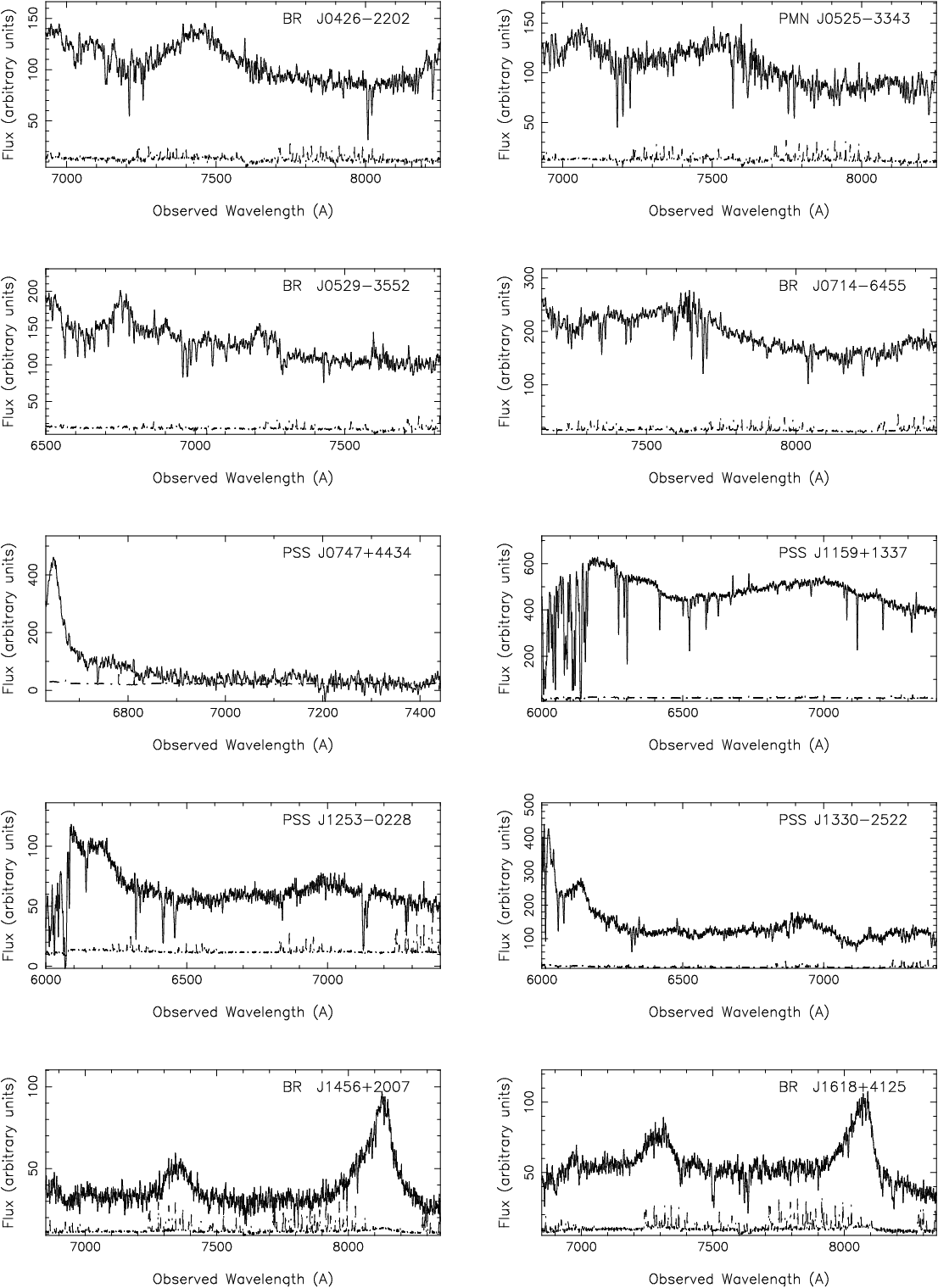}
\setcounter{figure}{0}
\caption{{\it Continued.}}
\end{figure*}

\begin{figure*}[t]
\includegraphics[width=.97\hsize,height=2.01\vsize]{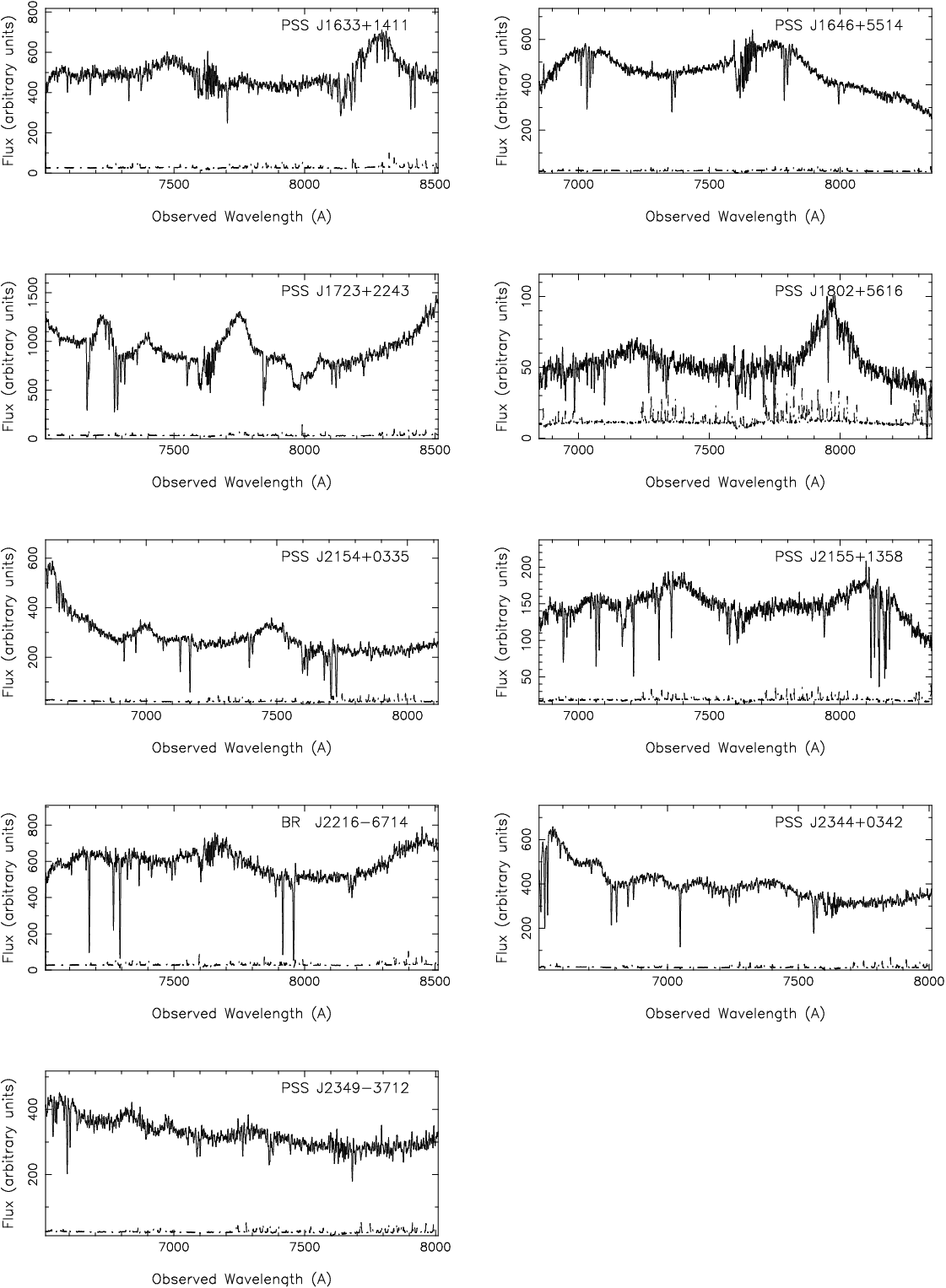}
\setcounter{figure}{0}
\caption{{\it Continued.}}
\end{figure*}

\section{Metal Identification}

\begin{figure*}[t]
  \begin{center}
    \includegraphics[width=.7\hsize,height=.65\vsize,angle=0]{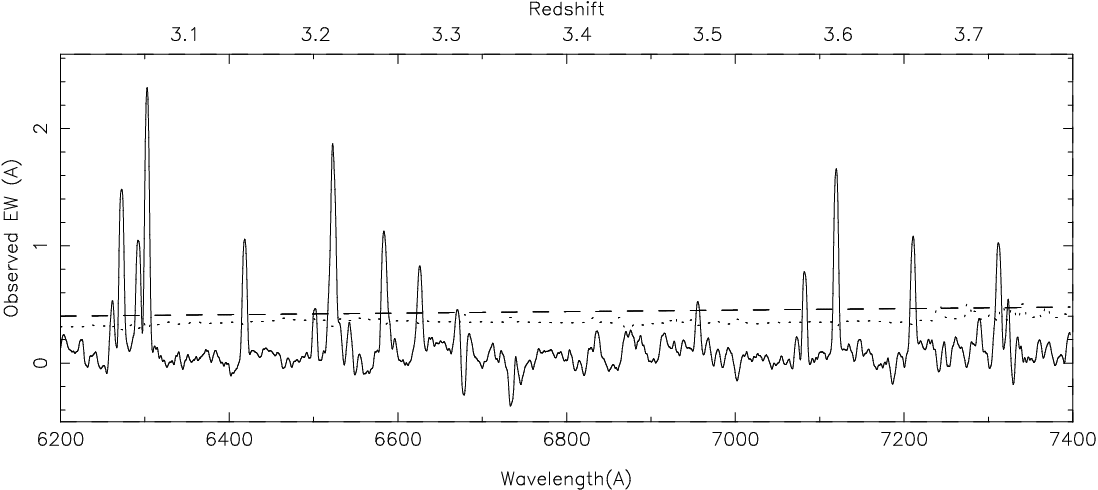}\hfill
    \caption{Example of the output of the automated metal line search
    (for quasar PSS J1159$+$1337). The solid line corresponds to the
    equivalent width spectrum and the dotted line to the error
    array. The straight dashed line represents the threshold above
    which the candidate absorbers are selected.  An example is the
    $z_{\rm abs}=3.724$ \civ\ doublet clearly visible at $\lambda_{\rm
    obs}\sim 7300$\AA\ and the corresponding \siiv\ doublet situated
    at $\lambda_{\rm obs}\sim 6600$\AA.}
\label{f:dz_ew} \end{center}
\end{figure*}

\subsection{Methodology}

To select metal lines we use a detection algorithm based on the method
presented in P\'eroux \etal\ (2001), supplemented by a visual
search. Following past work, the spectra are analysed from 5000 km
s$^{-1}$ blueward of the \civ\ emission line (to avoid lines possibly
associated with the quasar) towards shorter wavelengths. The analysis
is stopped when the signal-to-noise ratio becomes too low to detect a
\civ\ line at the 5$\sigma$ level (corresponding to $z_{min}$ in
Table~\ref{t:sn}). We then calculate the equivalent widths of all the
candidate lines.  Figure~\ref{f:dz_ew} shows an example (for PSS
J1159$+$1337) of the output of the algorithm we use to detect metal
lines - a box-like running average of the equivalent width as a
function of wavelength. In this figure, our detection threshold is
represented by the straight dashed line: any feature peaking above
this limit is a candidate absorber. This output is then visually
inspected to detect any lines which might have been missed by the
automatic procedure.  The lines are then identified by analysing each
spectrum in velocity space and the redshift of the system is measured
that best fits each line of a doublet and other associated lines. All
detected absorptions with confirmed identifications are listed in
Table~\ref{t:metal} of the appendix for each quasar spectrum.

\subsection{Notes on Individual Spectra}

\begin{enumerate}

\item BR  J0006$-$6208 ($z_{\rm em}=4.455$)  

$z_{\rm abs}=1.687$ -- Weak absorption but the wavelength alignment
and EW ratio are in excellent agreement with expectations from \mgii\
absorbers.

$z_{\rm abs}=1.956$ -- Symmetric profiles associated with a saturated
\mgii\ doublet. \mgi\ 2852 and many associate \feii\ lines (2374,
2382, 2586, 2600) are observed. Possible \fei\ 2719 and \sii\ 2515 are
also found at the same redshift.

$z_{\rm abs}=3.588$ -- Unambiguous two-component \civ\ doublet with
possible associate \ci\ 1560.


\item BR  J0030$-$5129 ($z_{\rm em}=4.174$)  

$z_{\rm abs}=3.465$ -- Narrow, well-defined \civ\ doublet.  


\item PSS J0034$+$1639 ($z_{\rm em}=4.293$) 

$z_{\rm abs}=1.588$ -- The two members of this \mgii\ doublet are not
exactly aligned, but EW ratio means that this is a definite
identification.

$z_{\rm abs}=1.798$ -- The 2796 line of this \mgii\ doublet is
saturated. Associated strong \feii\ 2382, 2586 and 2600 lines are
observed.

$z_{\rm abs}=1.883$ -- The 2803 line of this \mgii\ doublet is
slightly contaminated on its red side by \civ\ 1548 at $z_{\rm
abs}=4.225$.

$z_{\rm abs}=3.442$ -- Although the EW ratio is not precisely the one
expected from \civ\ doublets, the wavelength alignment and similarity
in profiles mean that this is most probably a \civ\ absorption. \ci\
1560 coincides exactly with the \civ\ 1548 at $z_{\rm abs}=3.478$.

$z_{\rm abs}=3.478$ -- Clean and unambiguous \civ\ doublet with
associated \siii\ 1526.

$z_{\rm abs}=4.225$ -- At red end of the spectrum, another \civ\
doublet is clearly detected. \ci\ 1328 coincides exactly with the
\civ\ 1550 at $z_{\rm abs}=3.478$ and \siii\ 1304 is detected.


\item PSS J0133$+$0400 ($z_{\rm em}=4.154$)  

This spectrum is particularly rich in \siiv\ doublets.

$z_{\rm abs}=1.664$ -- Saturated \mgii\ doublet with associated \mgi\
2852 absorption line. \feii\ 2586 and 2600 are also observed at this
redshift.

$z_{\rm abs}=3.138$ -- Although contamination between the two lines of
this \civ\ doublet means that the EW ratio is slightly off the
expected value, this is most probably a \civ\ system.

$z_{\rm abs}=3.229$ -- Small, well-defined \civ\ doublet.

$z_{\rm abs}=3.620$ -- This \siiv\ doublet might have associated
\civ\ absorption, but given the sky contamination at the relevant
wavelengths (around 7150 \AA), this cannot be confirmed.

$z_{\rm abs}=3.771$ -- This \civ\ doublet is characterised by an
asymmetric profile which is also seen in \siiv. A strong \cii\ 1334 is
observed at the blue end of the spectrum and additional \siii\ 1526
and \feii\ 1608 are detected at the same redshift. P\'eroux \e\ (2001)
also report a DLA at this redshift.

$z_{\rm abs}=3.992$ -- \civ\ system with associated a \siiv\ doublet.

$z_{\rm abs}=3.996$ -- Situated only 300 km s$^{-1}$ away from the
previous system, this \civ\ doublet has associated with \oi\ 1302 and
\siii\ 1304 absorptions.

$z_{\rm abs}=4.110$ -- Narrow \siiv\ doublet with a possible \ci\ 1280
line associated and a definite \siii\ 1304 feature (\ci\ 1280 is
blended with \civ\ 1548 at $z_{\rm abs}=3.229$).

$z_{\rm abs}=4.115$ -- Situated only 300 km s$^{-1}$ away from the
previous system, this is another narrow \siiv\ doublet with associated
\siii\ 1526.


\item PSS J0248$+$1802 ($z_{\rm em}=4.422$)  

$z_{\rm abs}=1.470$ -- Although \mgii\ 2796 is contaminated on its red
wing, this is most probably a \mgii\ doublet.

$z_{\rm abs}=3.969$ -- \siiv\ doublet with \siiv\ 1393 blended with
\mgii\ 2803 at $z_{\rm abs}=1.470$.

$z_{\rm abs}=4.185$ -- Another medium-strength \siiv\ doublet.


\item BR  J0311$-$1722 ($z_{\rm em}=4.039$)  

$z_{\rm abs}=4.034$ -- This definite \civ\ doublet has an asymmetric
profile. \oi\ 1302 is possibly detected at the same redshift.


\item BR  J0324$-$2918 ($z_{\rm em}=4.622$)  

$z_{\rm abs}=1.923$ -- Weak \mgii\ doublet.

$z_{\rm abs}=2.227$ -- Many \feii\ lines (2344, 2382, 2586 and 2600)
are detected at this redshift, but our spectrum does not cover the
appropriate wavelength range for the detection of the \mgii\ doublet.

$z_{\rm abs}=3.746$ -- Weak \civ\ doublet.


\item BR  J0334$-$1612 ($z_{\rm em}=4.363$)  

$z_{\rm abs}=3.668$ -- Narrow \civ\ doublet.

$z_{\rm abs}=3.759$ -- Another narrow \civ\ doublet.

$z_{\rm abs}=3.788$ -- A \siiv\ doublet is detected at this wavelength.

$z_{\rm abs}=4.186$ -- Another \siiv\ doublet is detected, with the
\siiv\ 1393 line blended with \civ\ 1548 at $z_{\rm abs}=3.668$.


\item BR  J0355$-$3811 ($z_{\rm em}=4.545$)  

$z_{\rm abs}=1.986$ -- This saturated \mgii\ doublet shows clear
evidences of a two-component profile. Associated \feii\ 2586 and 2600
lines are observed.

$z_{\rm abs}=3.855$ -- Unusually broad and weak \civ\ doublet.

$z_{\rm abs}=4.313$ -- Weak \civ\ doublet.

$z_{\rm abs}=4.318$ -- Situated 300 km s$^{-1}$ away from the
previous system, another weak \civ\ system.


\item BR  J0419$-$5716 ($z_{\rm em}=4.461$)  

$z_{\rm abs}=2.028$ -- Many \feii\ lines (2382, 2586 and 2600)
are detected at this redshift, but our spectrum does not cover the
appropriate wavelength range for the detection of the \mgii\ doublet.

$z_{\rm abs}=4.218$ -- The \civ\ 1550 line of this doublet is probably
contaminated by another absorption, which explains the differences
in the profile shape of this \civ\ doublet.

$z_{\rm abs}=4.311$ -- Unambiguous \civ\ doublet.

$z_{\rm abs}=4.434$ -- Weak \civ\ doublet where the \civ\ 1548 line is
blended with a stronger unidentified absorption.


\item BR  J0426$-$2202 ($z_{\rm em}=4.320$)  

$z_{\rm abs}=3.606 $ -- This \civ\ doublet shows a characteristic
two-component profile.

$z_{\rm abs}=3.682$ -- Weak \civ\ doublet.

$z_{\rm abs}=4.172$ -- Nicely defined \civ\ doublet with associated
\siiv\ absorption.

\item PMN J0525$-$3343 ($z_{\rm em}=4.383$)  

$z_{\rm abs}=1.568$ -- Well defined \mgii\ doublet. 

$z_{\rm abs}=3.581$ -- Weak \civ\ doublet with associated \feii\ 1608
absorption.

$z_{\rm abs}=4.063$ -- Weak \civ\ doublet with associated \siiv\
doublet (the \siiv\ 1402 is blended with \civ\ 1548 at $z_{\rm
abs}=3.581$), \siii\ 1526 and \feii\ 1608.

$z_{\rm abs}=4.313$ -- Slightly broad \civ\ doublet.

$z_{\rm abs}=4.430$ -- \siiv\ doublet.

\item BR  J0529$-$3552 ($z_{\rm em}=4.172$)  

$z_{\rm abs}=1.398$ -- Fairly weak \mgii\ doublet.

$z_{\rm abs}=1.423$ -- The profile of this \mgii\ doublet is
characterised by a square shape.

$z_{\rm abs}=1.656$ -- Another \mgii\ doublet is observed at this
redshift.

$z_{\rm abs}=3.503$ -- Well defined \civ\ doublet with associate
\feii\ 1608 and \alii\ 1670 absorption lines.

$z_{\rm abs}=4.062$ -- This \siiv\ doublet is accompanied with \siii\
1304 \& 1526 and \cii\ 1304.

\item BR  J0714$-$6455 ($z_{\rm em}=4.462$)  

$z_{\rm abs}=3.747$ -- This \civ\ doublet is associated with a \siii\
1526 line. 

$z_{\rm abs}=3.966$ -- Another well defined \civ\ doublet is detected
at this redshift.

$z_{\rm abs}=4.193$ -- Both \civ\ and \siiv\ doublets are observed,
but \siiv\ 1402 is blended with \siii\ 1526 at $z_{\rm abs}=3.747$.

\item PSS J0747$+$4434 ($z_{\rm em}=4.430$)  

This spectrum is of very low quality and no lines could be
identified.

\item PSS J1159$+$1337 ($z_{\rm em}=4.073$)  

$z_{\rm abs}=1.739$ -- Many \feii\ lines (2344, 2374, 2382, 2586 and
2600) are detected at this redshift, but our spectrum does not cover
the appropriate wavelength range for the detection of the \mgii\
doublet.

$z_{\rm abs}=3.724$ -- \civ\ and \siiv\ doublets are detected at this
redshift (the \siiv\ line is blended with \siiv\ 1393 at
3.756). P\'eroux \e\ (2001) also report a DLA at this redshift.

$z_{\rm abs}=3.756$ -- Another \siiv\ doublet is observed.

\item PSS J1253$-$0228 ($z_{\rm em}=4.007$)  

$z_{\rm abs}=1.261$ -- A strong \mgii\ doublet is detected at this
redshift.

$z_{\rm abs}=3.606$ -- This \civ\ doublet also shows strong \siiv\
doublet and \cii\ 1334 line with the same characteristic two component
profile. P\'eroux \e\ (2001) also report a sub-DLA (see P\'eroux \e\
2003 for a definition of this class of systems) at this redshift.

\item PSS J1330$-$2522 ($z_{\rm em}=3.949$)  

$z_{\rm abs}=3.084$ -- Associated with this \civ\ doublet is \alii\
1670. P\'eroux \e\ (2001) also report a sub-DLA (see P\'eroux \e\
2003) at this redshift.

$z_{\rm abs}=3.390$ -- Another \civ\ doublet is observed at this
redshift.

$z_{\rm abs}=3.772$ -- A weak \siiv\ doubled is detected.

$z_{\rm abs}=3.769$ -- At the red end of the spectrum, another \civ\
doublet is detected.

\item BR  J1456$+$2007 ($z_{\rm em}=4.249$)  

$z_{\rm abs}=1.761$ -- Narrow \mgii\ doublet.

$z_{\rm abs}=4.036$ -- The \civ\ 1550 line of this \civ\ doublet is
slightly contaminated.

\item BR  J1618$+$4125 ($z_{\rm em}=4.213$)  

$z_{\rm abs}=4.223$ -- This \siiv\ system is the only doublet which
can be convincely identified.

\item PSS J1633$+$1411 ($z_{\rm em}=4.351$)  

$z_{\rm abs}=3.580$ -- Nice \civ\ doublet

$z_{\rm abs}=4.150$ -- A \siiv\ doublet is identified at this
redshift.

$z_{\rm abs}=4.282$ -- This \civ\ doublet has a characteristic profile
which includes a smaller component to the blue of the main feature.

\item PSS J1646$+$5514 ($z_{\rm em}=4.037$)

This spectrum is particularly rich in \civ\ systems which are visible as
clustered pairs of doublets.

$z_{\rm abs}=1.859$ -- Well defined \mgii\ doublet where the EW ratio
is in excellent agreement with expectations from physical properties.

$z_{\rm abs}=3.516$ -- A weak \civ\ doublet is detected at this
redshift.

$z_{\rm abs}=3.521$ -- This medium strength \civ\ system has its \civ\ 1548
line blended with the \civ\ 1550 line at $z_{\rm abs}=3.544$.

$z_{\rm abs}=3.544$ -- In the close vicinity of the previous system is
another stronger \civ\ doublet.

$z_{\rm abs}=3.754$ -- This \civ\ doublet has its lines separated by
the \civ\ 1548 line at $z_{\rm abs}=3.759$.

$z_{\rm abs}=3.759$ -- This weaker \civ\ doublet is crossing the
previous doublet but can nevertheless be fairly well isolated for EW
measurements.

$z_{\rm abs}=4.030$ -- Similarly to the two doublets described above,
this \civ\ system is crossed over by the $z_{\rm abs}=4.037$ \civ\
absorber, although the closer proximity between the two means that the
EW are not as reliable. The \siiv\ doublet, \siii\ 1526 and \feii\
1608 are also detected at this redshift.

$z_{\rm abs}=4.037$ -- Weaker \civ\ absorber as described above.

\item PSS J1723$+$2243 ($z_{\rm em}=4.520$) 

$z_{\rm abs}=3.696$ -- Characteristic two-component \siii\ 1526,
\feii\ 1608 and \alii\ 1670 are detected at this redshift. The shape
of features means that the identification is unambiguous. Nevertheless
the corresponding \civ\ doublet is not convincing due to an odd EW
ratio.

$z_{\rm abs}=3.704$ -- Well defined and strong \civ\ doublet.

$z_{\rm abs}=4.234$ -- This \civ\ doublet is blended in its \civ\ 1550
line with the \civ\ 1548 at $z_{\rm abs}=4.244$.

$z_{\rm abs}=4.244$ -- The \civ\ system identified at this redshift is
clustered with the doublet described previously. The consequences of
this arrangement is that the EW measurements are less reliable. \siiv\
is nevertheless observed at this redshift.

\item PSS J1802$+$5616 ($z_{\rm em}=4.158$)  

$z_{\rm abs}=4.048$ -- A \siiv\ doublet is identified at this redshift.

$z_{\rm abs}=4.190$ -- A weak \civ\ doublet is detected at this
redshift. The \civ\ 1548 line is probably contaminated to some extend.

\item PSS J2154$+$0335 ($z_{\rm em}=4.363$)  

$z_{\rm abs}=1.756$ -- This saturated \mgii\ absorber is associated
with a \mgi\ 2852 absorption. \feii\ 2586 and 2600 are also detected
at this redshift.

$z_{\rm abs}=3.776$ -- The good alignment and EW ratio means that
this a definite \civ\ doublet. Nevertheless, the \civ\ 1550 line is
clearly contaminated in its blue wing by some unidentified absorption
which means that the EW measurement of this line is not
reliable. P\'eroux \e\ (2001) also report a sub-DLA (see P\'eroux \e\
2003) at this redshift.

$z_{\rm abs}=3.961$ -- Very well defined \civ\ doublet is observed at
this redshift. The associated \siiv\ doublet is also detected.

\item PSS J2155$+$1358 ($z_{\rm em}=4.256$)  

$z_{\rm abs}=1.914$ -- The characteristic two-component saturated
profile is unambiguously associated with a \mgii\ doublet. \feii\
(2382, 2586 and 2600) lines are also detected at this redshift.

$z_{\rm abs}=3.567$ -- Although the EW ratio of this doublet is not
in excellent agreement with expectations from \civ\ physical
properties, we identify this system as a \civ\ absorber.

$z_{\rm abs}=4.243$ -- This \civ\ doublet is saturated which means
that the EW ratio is not reliable. The identification is nevertheless
reliable. A strong \siiv\ doublet is also detected at this redshift.

\item BR  J2216$-$6714 ($z_{\rm em}=4.469$)  

$z_{\rm abs}=2.060$ -- Many \feii\ lines (2344, 2374, 2382, 2586 and
2600) are detected at this redshift, but our spectrum does not cover
the appropriate wavelength range for the detection of the \mgii\
doublet.

$z_{\rm abs}=3.728$ -- Although the profiles are quite different for
the two lines of this \civ\ doublet, the identification is secure
due to a good alignment in velocity space and an EW ratio that conforms to
expectations from the physical properties of these systems.

$z_{\rm abs}=3.837$ -- Another \civ\ doublet is observed at this
wavelength.

$z_{\rm abs}=4.095$ -- A weak \civ\ system is present at this redshift.

\item PSS J2344$+$0342 ($z_{\rm em}=4.239$)  

$z_{\rm abs}=3.219$ -- This \civ\ doublet is probably saturated which
would explain the unexpected EW ratio. \feii\ 1608 and \alii\ 1670 are
also detected at this redshift. P\'eroux \e\ (2001) also report a DLA
at this redshift.

$z_{\rm abs}=4.053$ -- There is a weak \civ\ doublet at this redshift
and associated \siii\ 13094 and 1526 absorption lines.

$z_{\rm abs}=3.882$ -- Both \civ\ and \siiv\ are clearly observed at this redshift.

\item BR  J2349$-$3712 ($z_{\rm em}=4.208$)  

$z_{\rm abs}=1.758$ -- Although the \mgii\ 2803 line of this doublet
is slightly contaminated, this system is most probably a weak \mgii\
absorber.

$z_{\rm abs}=3.258$ -- A \civ\ doublet is clearly identified at this
redshift.

$z_{\rm abs}=3.284$ -- A weak \civ\ system is detected at this
redshift. \feii\ 1608, \alii\ 1670 and \siii\ 1526 are also observed
at the same redshift.

$z_{\rm abs}=3.690$ -- Yet another \civ\ doublet is detected at this
redshift. The \siiv\ doublet and \alii\ 1670 are also detected at this
redshift. The \siiv\ 1393 line is blended with \siii\ 1526 at $z_{\rm
abs}=3.284$. P\'eroux \e\ (2001) also report a sub-DLA (see P\'eroux
\e\ 2003) at this redshift.

$z_{\rm abs}=3.756$ -- A weak \civ\ absorber is found at this
redshift, although the profiles of both lines are not totally
identical.

$z_{\rm abs}=3.960$ -- A nice \civ\ system is detected. It is
characterised by a two-component profile.


\end{enumerate}

\begin{figure}[t]
  \begin{center}
    \includegraphics[width=.25\vsize,height=.9\hsize,angle=-90]{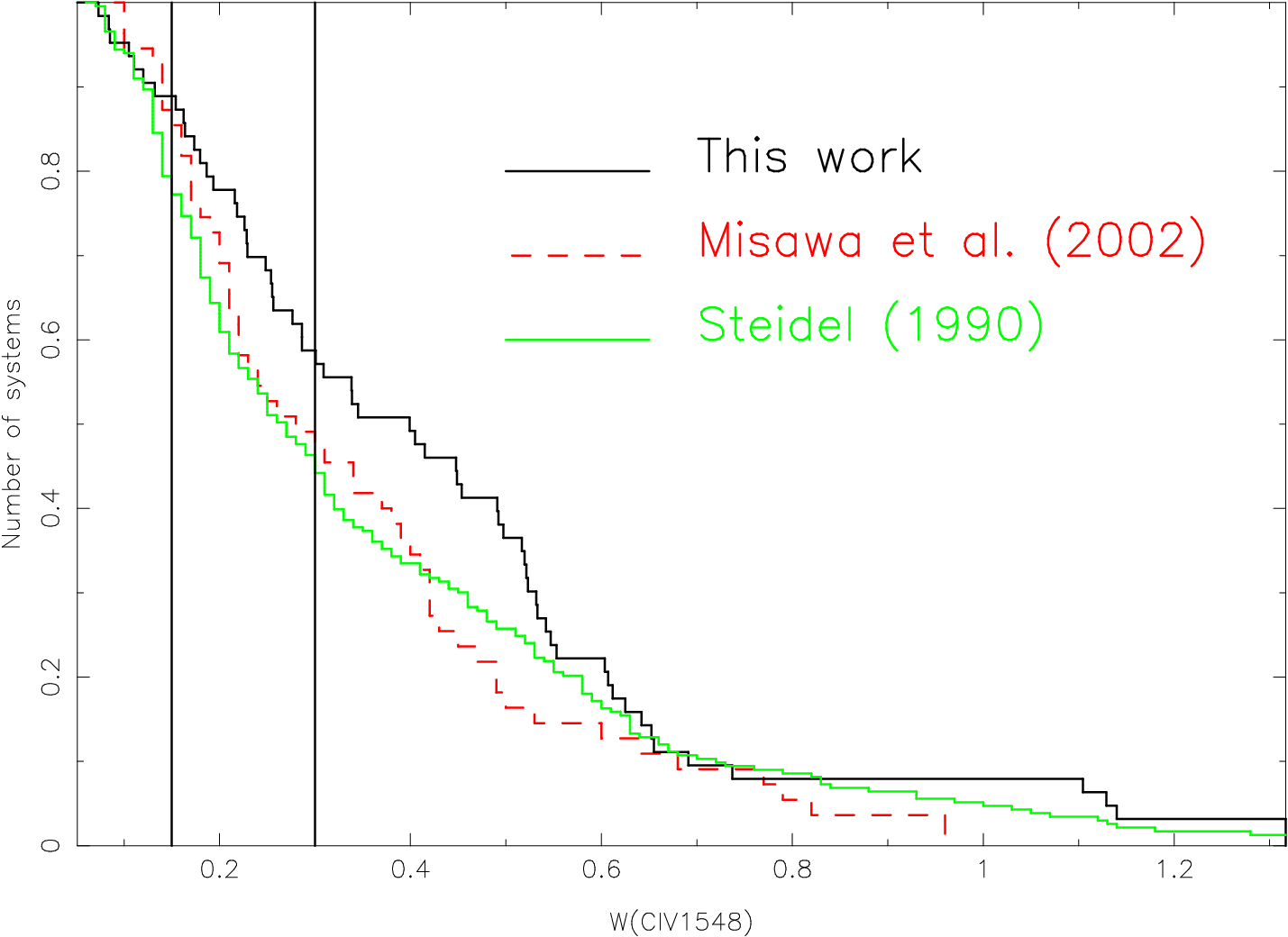}\hfill
    \includegraphics[width=.25\vsize,height=.9\hsize,angle=-90]{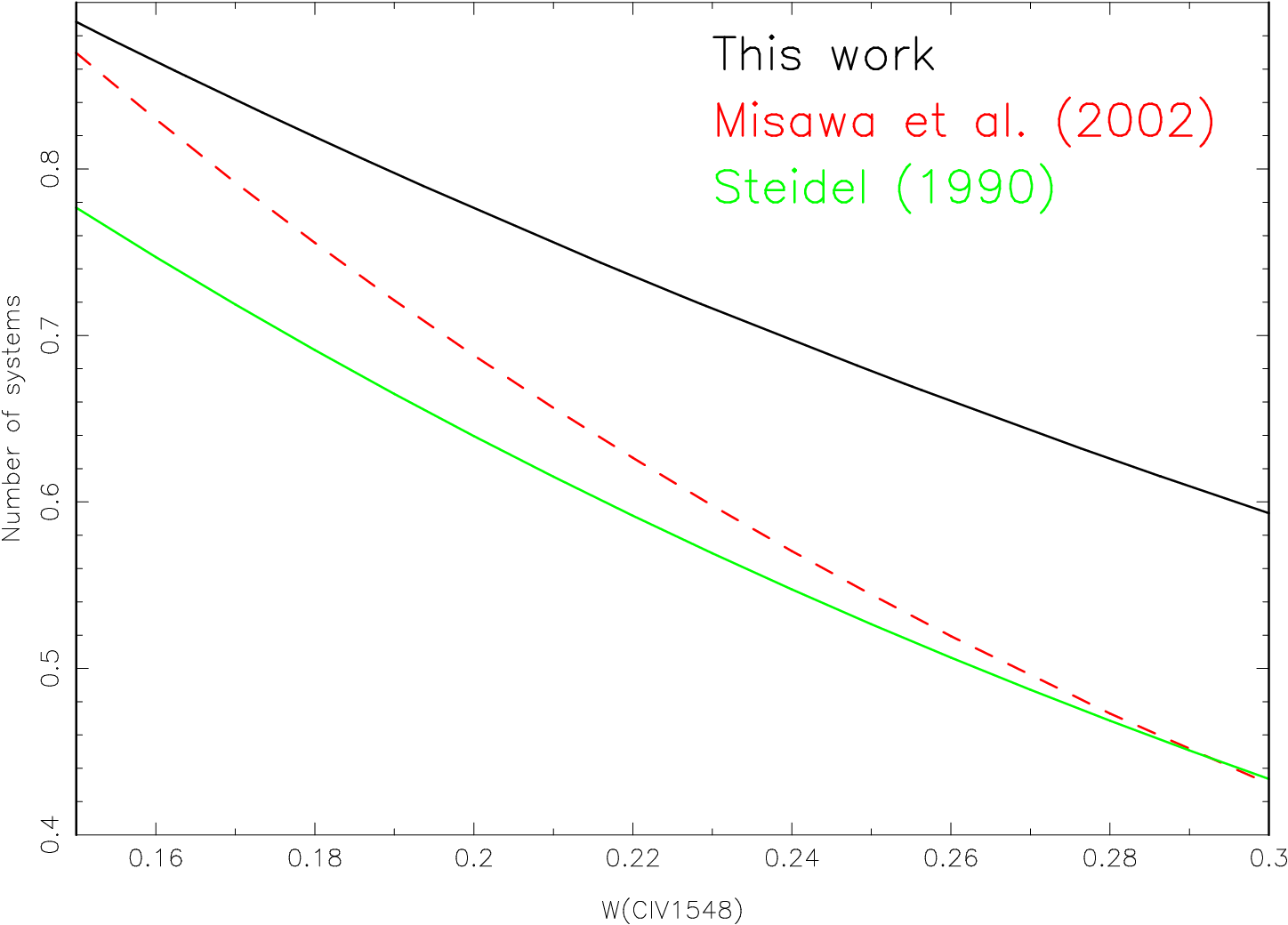}\hfill
    \caption{{\bf Top panel:} Normalised cumulative distributions of
    \civ\ doublets as a function of their equivalent widths at rest,
    $W_{\rm rest}$. The black line is for this work, the dashed
    light-coloured line for the Misawa et al. (2002) sample and the solid
    light-coloured line for Steidel (1990). The change of slope of the
    Misawa et al. distribution at $0.15<W_{\rm rest}<0.30$ with
    respect to the two other samples, show that the latter are not
    complete at this equivalent width cut-off. Indeed, the study of
    Misawa et al. (2002) is based on higher-resolution data, and
    therefore their sample is more complete than the previous work of
    Steidel (1990) explaining why results from the two surveys
    differ. {\bf Bottom panel:} Exponential fits to the cumulative
    distributions in the region of interest. The steeper slope of the
    Misawa et al. (2002) sample is clearly visible.} \label{f:cumu}
    \end{center}
\end{figure}

\begin{table}[!h]
\begin{center}
\begin{tabular}{lcccc}
\hline \hline 
Absorber &  all&  $W_{\rm rest}>0.15$&$W_{\rm rest}>0.30$&$W_{\rm rest}>0.70$    \\
\hline 
\civ\    &64   &41   &19   &...\\
\mgii\   &19   &14   &11   &7  \\
\siiv\   &28   &14   &3    &...\\
\hline\hline
\end{tabular}
\end{center}
\caption{Number of absorber for a given $W_{\rm rest}$ limit.}
\label{t:nber}
\end{table}

The whole sample is composed of 64 \civ\, 19 \mgii\ and 28 \siiv\
doublets. For comparison Steidel (1990) use 11 spectra at $3.04<z_{\rm
em}<4.11$ in addition to the 55 spectra at $1.08<z_{\rm em}<3.56$ from
Sargent, Boksenberg \& Steidel (1988) and finds a total of 275
\civ. Steidel \& Sargent (1992) find 111 \mgii\ absorbers in 103
quasar spectra. Songaila (2001) has looked for \civ and \siiv\
doublets in the spectra of 32 quasars with $2.31<z_{\rm em}<5.86$ down
to $\log (N_{\rm CIV})>12.0$. She finds a total of 139 \civ\ lines and
49 \siiv\ lines at $z_{\rm abs} > 3.5$, with 122 and 39, respectively
for $3.5 < z_{\rm abs} < 4.5$, roughly the redshift limits of this
survey. More recently, Misawa et al. (2002) find 55 \civ\ doublets, 19
\siiv\ and only 3 \mgii\ in the spectra of 18 quasars. The \civ\
samples of each of these surveys are plotted in Figure~\ref{f:cumu} as
a function of the \civ\ $\lambda$ 1548 rest frame equivalent widths,
$W_{\rm rest}$. The change of slope of the Misawa et al. distribution
at $0.15<W_{\rm rest}<0.30$\AA\ conforms to theoretical expectations
of the evolution of quasar absorbers (Churchill \& Vogt 2001). On the
other hand, the fact that the other samples do {\it not} show such
strong evolution indicates that they are not complete below $W_{\rm
rest}<0.30$\AA. Indeed, the study of Misawa et al. (2002) is based on
higher-resolution data, and therefore their sample is more complete
than the previous work of Steidel (1990) explaining while results from
the two surveys differ.  Since our sample suffers the same problem of
incompleteness as Steidel (1990), we choose to limit our analysis to
doublets with $W_{\rm rest}>0.30$\AA. For a Doppler parameter $b<20$,
this corresponds to $\log$ N(\civ) $>14.5$ (Steidel et al., 1990). For
reference, Table~\ref{t:nber} summarises the number of absorbers in
various sub-samples. Interestingly, we find proportionally more \mgii\
doublets than Misawa et al. (2002) but less \siiv\ systems, and
therefore do not consider the latter absorber in our statistical
analysis. We find four $W_{\rm rest}>1$\AA\ \civ\ absorber in our
sample which demonstrates that such systems exist at $z>3$.

\section{Statistical Properties}

\subsection{Number Density}

\begin{figure}[t]
  \begin{center}
    \includegraphics[width=0.75\vsize,height=.9\hsize,angle=-90]{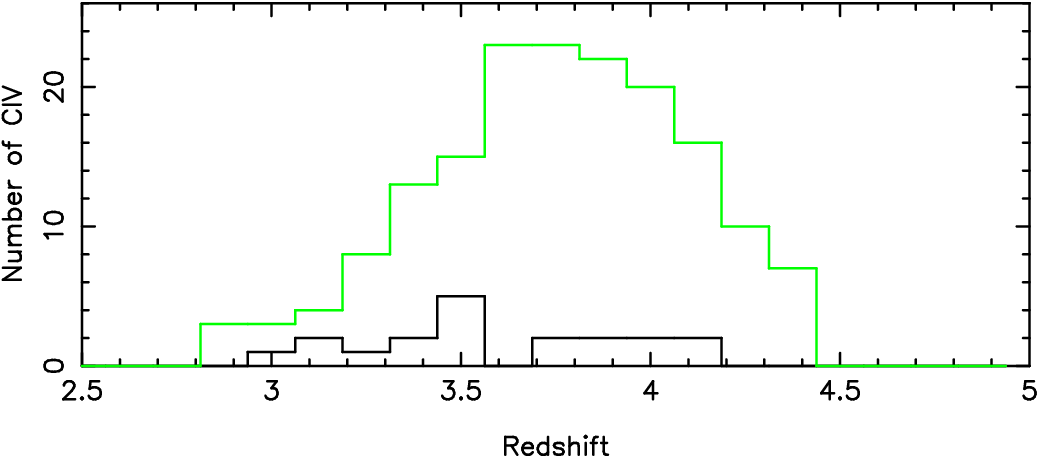}\hfill
    \includegraphics[width=0.75\vsize,height=.9\hsize,angle=-90]{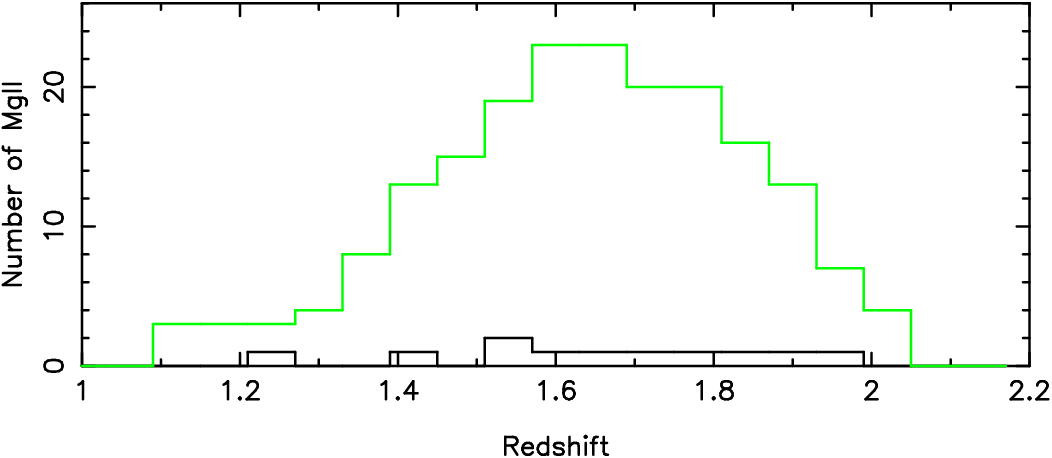}\hfill
    \caption{The light-coloured histograms represent the survey
sensitivity, $g(z)$, i.e. the cumulative number of lines of sight
along which a $W_{\rm rest}>0.30$ \AA\ \civ\ (top panel) or a \mgii\
(bottom panel) {\it could} be detected at the 5$\sigma$ confidence
level. The black histograms are the redshift distribution of the
sub-samples of $W_{\rm rest}>0.30$ \AA\ \civ\ and \mgii,
respectively.}
    \label{f:gz}
  \end{center}
\end{figure}

\begin{table}[ht]
\begin{center}
\begin{tabular}{lccc}
\hline \hline 
Quasar &    S/N&$z_{\rm min}$ &$z_{\rm max}$  \\
\hline 
BR  J0006$-$6208   &20$-$30 &3.531   &4.446  \\
BR  J0030$-$5129   &10$-$20 &3.208   &4.165  \\
PSS J0034$+$1639   &15$-$25 &3.274   &4.243  \\
PSS J0133$+$0400   &30$-$40 &3.115   &4.077  \\
PSS J0248$+$1802   &10$-$20 &3.287   &3.807$^*$  \\
BR  J0311$-$1722   &10$-$20 &3.206   &4.031  \\
BR  J0324$-$2918   &20$-$30 &3.621   &4.471  \\
BR  J0334$-$1612   &10$-$20 &3.287   &3.416$^*$  \\
		   &        &3.423   &3.802$^*$  \\
BR  J0355$-$3811   &20$-$30 &3.621   &4.471  \\
BR  J0419$-$5716   &10$-$20 &3.621   &4.452  \\
BR  J0426$-$2202   &10$-$20 &3.479   &4.311  \\
PMN J0525$-$3343   &10$-$20 &3.479   &4.329  \\
BR  J0529$-$3552   &20$-$30 &3.201   &4.051  \\
BR  J0714$-$6455   &20$-$30 &3.621   &4.453  \\
PSS J0747$+$4434   &5$-$10  &3.287   &3.807$^*$  \\
PSS J1159$+$1337   &40$-$50 &2.788   &3.889  \\
PSS J1253$-$0228   &15$-$25 &2.811   &3.857  \\
PSS J1330$-$2522   &10$-$20 &2.807   &3.887  \\
BR  J1456$+$2007   &5$-$10  &3.366   &4.240$^*$  \\
BR  J1618$+$4125   &10$-$20 &3.352   &4.204  \\
PSS J1633$+$1411   &20$-$30 &3.531   &4.342  \\
PSS J1646$+$5514   &20$-$30 &3.343   &4.029  \\
PSS J1723$+$2243   &15$-$25 &3.530   &4.158  \\
		   &        &4.166   &4.498  \\
PSS J1802$+$5616   &5$-$15  &3.351   &4.149$^*$  \\
PSS J2154$+$0335   &20$-$30 &3.276   &4.243  \\
PSS J2155$+$1358   &20$-$30 &3.368   &4.247  \\
BR  J2216$-$6714   &20$-$30 &3.531   &4.460  \\
PSS J2344$+$0342   &20$-$30 &3.209   &4.174  \\
BR  J2349$-$3712   &20$-$30 &3.209   &4.174  \\
dz({\rm total})	   &	    &&24.98\\	
dz($W_{\rm rest}>0.30$\AA)&  &&21.76\\	
\hline\hline
\end{tabular}
\end{center}
\scriptsize{$^*$ This object was not included into the final $W_{\rm
rest}>0.30$\AA\ redshift path.}\\
\caption{S/N ratios and redshift path surveyed for each spectrum.}
\label{t:sn}
\end{table}

In order to calculate the number density, $n(z)$, of absorbers, one
needs to measure the redshift path surveyed by the quasar sample.  For
each quasar spectrum, we compute the theoretical signal-to-noise (S/N)
necessary to recover an absorber with $W_{\rm rest}>0.30$\AA\ at
5$\sigma$. After measuring the S/N in each observed spectrum,
we reject 5 lines of sight: PSS J0248$-$1722, BR J0334$-$1612 and PSS
J0747$+$4434, all observed at WHT in November 2001 when the weather
conditions were extremely poor and BR J1456$+$2007 $\&$ PSS
J1802$+$5616 (WHT run of June 2002). The two last columns of
Table~\ref{t:sn} give the minimum and maximum redshift for each quasar
spectrum. This is illustrated by Figure~\ref{f:gz} which shows the
cumulative number of lines of sight along which a $W_{\rm
rest}>0.30$\AA\ \civ\ (top panel) or a \mgii\ (bottom panel) {\it
could} be detected at the 5$\sigma$ confidence level. This is the
survey sensitivity, $g(z)$, defined as follows,

\begin{equation}
g(z) = \sum H (z^{max}_{i} - z) H (z - z^{min}_{i})
\end{equation}

where H is the Heaviside step function. On the same figure the
redshift distribution of the sub-samples of $W_{\rm rest}>0.30$\AA\
\civ\ and \mgii, respectively, are plotted.

The number density of quasar absorbers is the number of absorbers,
$n$, per unit redshift $dz$, i.e., $dn/dz = n(z)$. This is a directly
observable quantity, although, its interpretation is dependent on the
geometry of the Universe. Indeed, the evolution of the number density
of absorbers with redshift is the intrinsic evolution of the true
number of absorbers combined with effects due to the expansion of the
Universe. For a non-zero $\Lambda$-Universe, the expected incidence of
absorbers, under the null hypothesis that there has been no evolution in
number density, gas cross-section, kinematics or chemical content, can
be described as (Tytler 1981)

\begin{equation}
n(z)=n_0 (1+z)^2 \times
(\Omega_0(1+z)^3+\Omega_k(1+z)^2+\Omega_{\Lambda})^{-1/2}
\end{equation}

For a flat $\Omega_{\Lambda}=0.7$, $\Omega_0=0.3$ Universe, a
non-evolving number density can be modelled with the following:

\begin{equation}
n(z)=n_0 \times (1+z)^2 \times
(0.3 \times (1+z)^3+ 0.7)^{-1/2}
\end{equation}

\begin{figure}[t]
  \begin{center}
    \includegraphics[width=.75\vsize,height=.9\hsize,angle=-90]{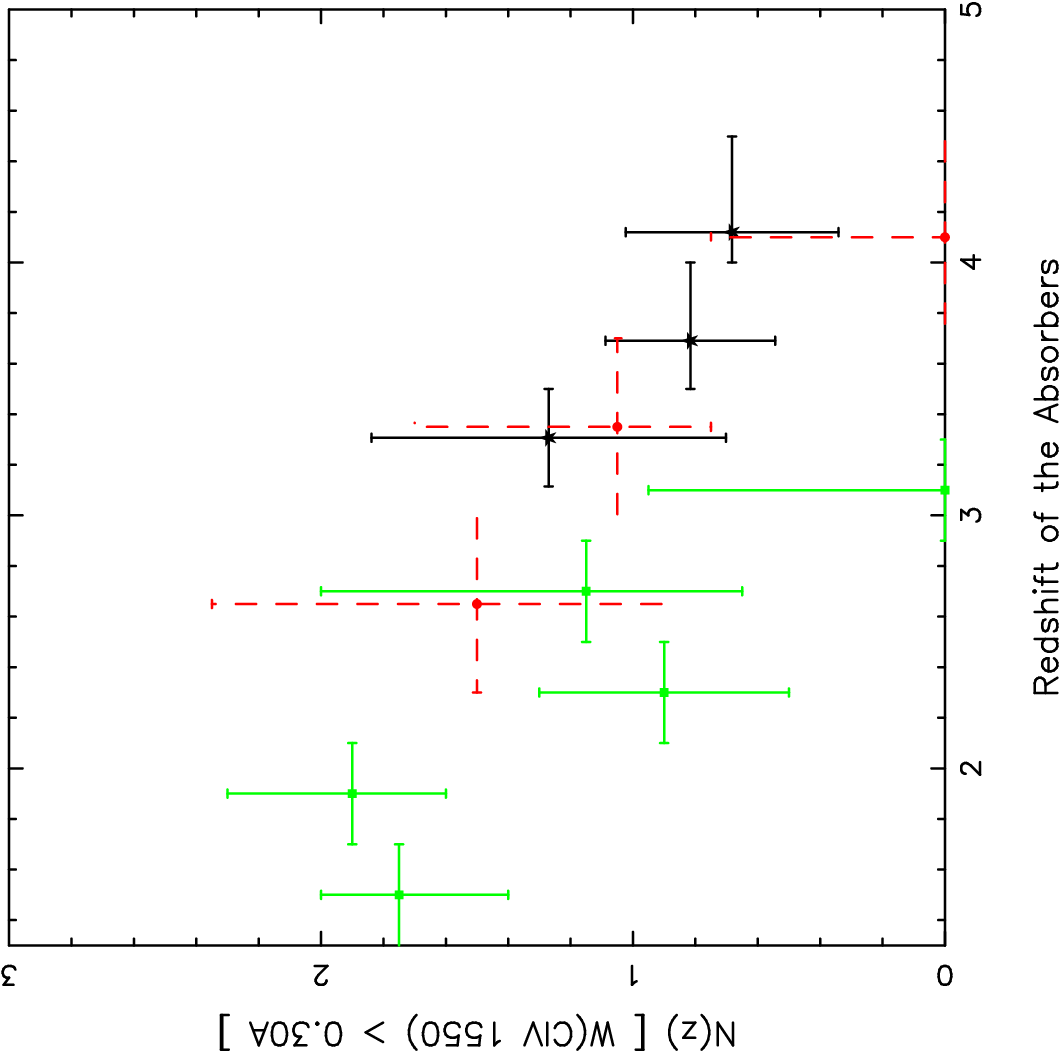}\hfill
    \includegraphics[width=.75\vsize,height=.9\hsize,angle=-90]{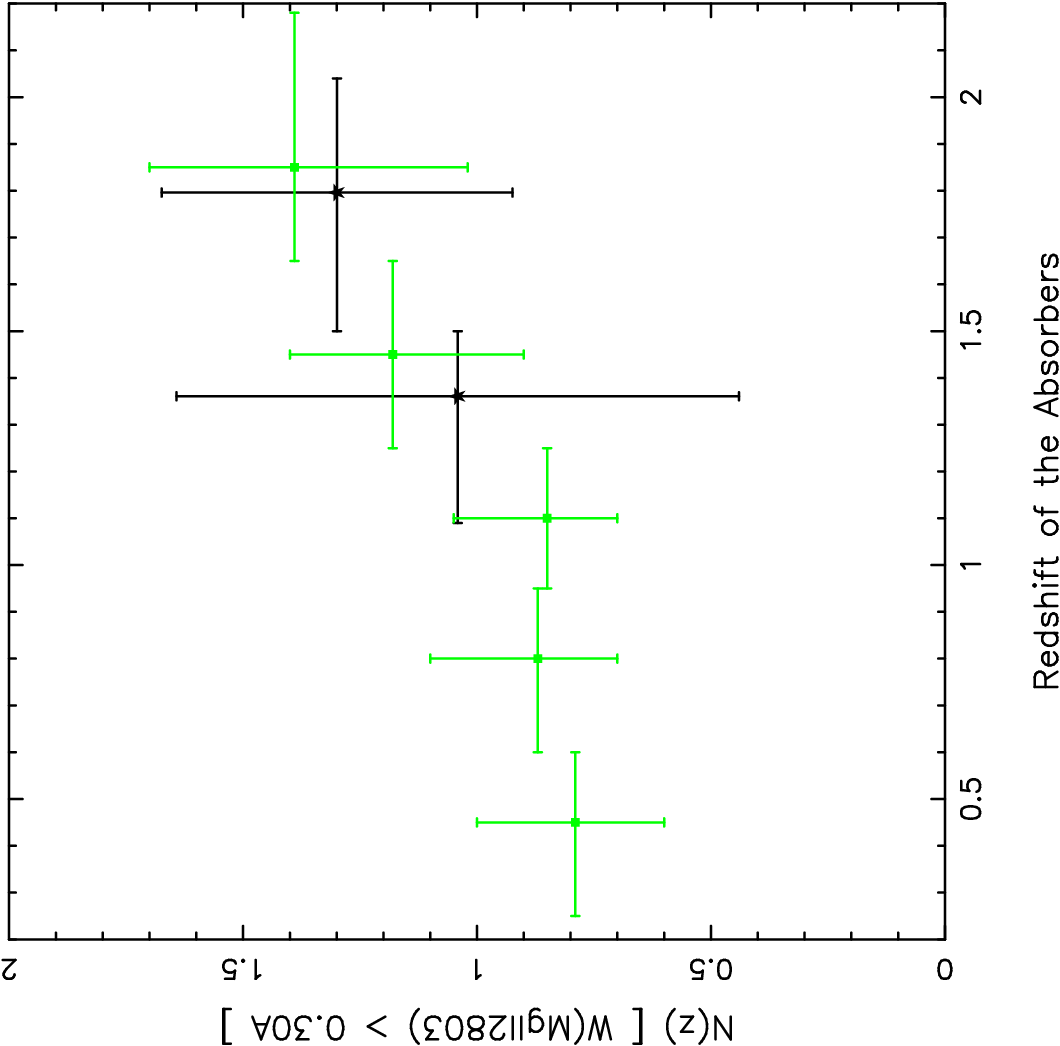}\hfill
    \caption{Number density of \civ\ doublets (top panel) and \mgii\
    systems (bottom panel) at $W_{\rm rest}>0.30$\AA.{\bf Top panel:}
    the high redshift bins are results from the present work (solid
    black), while Misawa et al. (2002) survey is at intermediate
    redshift (dashed grey) and Steidel (1990) at low redshift (solid
    light-coloured). {\bf Bottom panel:} The bins covering the all
    redshift range (light coloured) are from Steidel \& Sargent (1992)
    and the two bins at higher redshift are results from present work
    (in black). }
    \label{f:nz}
  \end{center}
\end{figure}

We compute the number density of $W_{\rm rest}>0.30$\AA\ \civ\ (top
panel of Figure~\ref{f:nz}) and compare our results with the previous
determinations of Steidel (1990) and Misawa et al. (2002) at lower
redshifts. It appears that in these two studies the last bin is an
underestimate of n(z) since no \civ\ doublets are detected in a range
where they might be expected to be seen. Most probably these are due
to the limitation of the surveys. Nevertheless, the combination of
these various studies show that the number density of $W_{\rm
rest}>0.30$\AA\ \civ\ absorbers decreases with redshift.

\begin{figure}[t]
  \begin{center}
    \includegraphics[width=.25\vsize,height=.9\hsize,angle=-90]{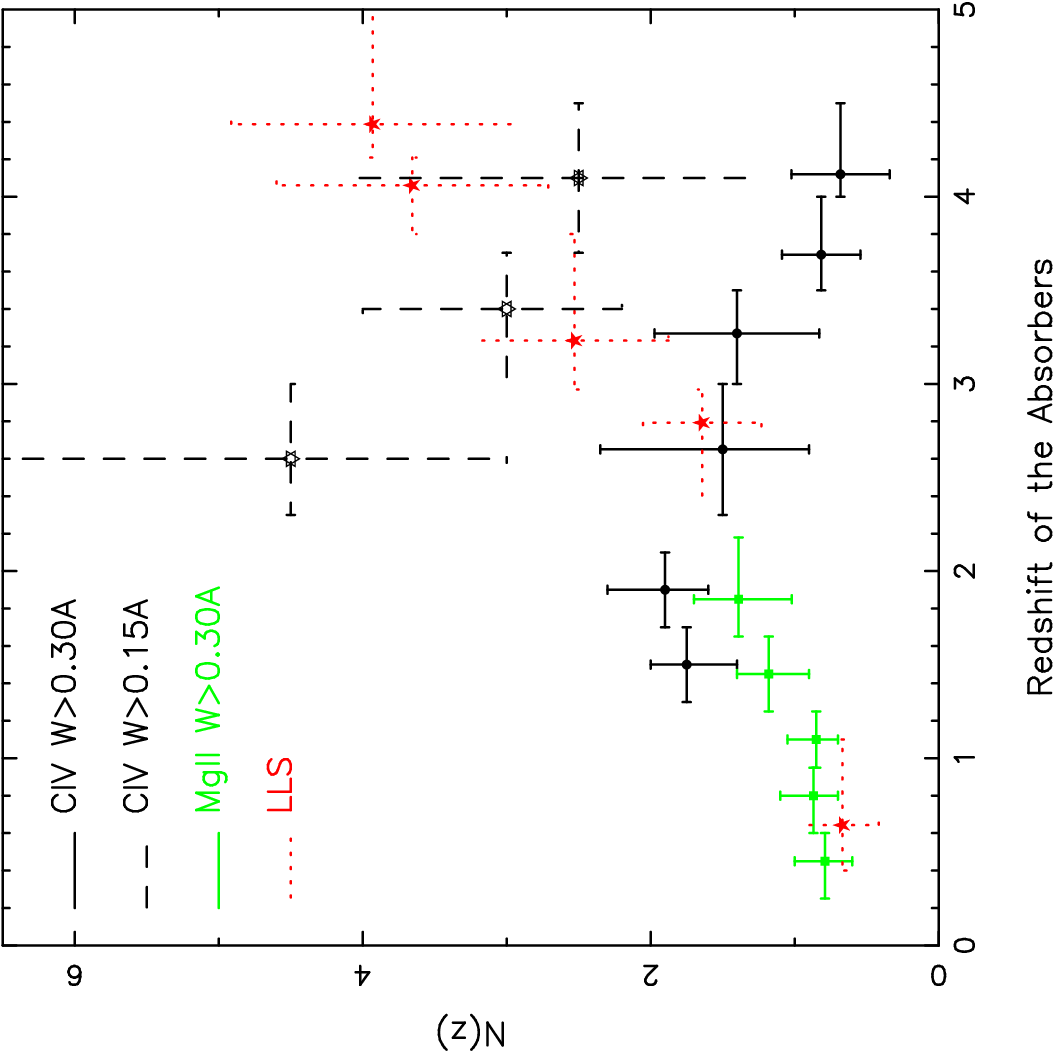}\hfill
    \includegraphics[width=.25\vsize,height=.9\hsize,angle=-90]{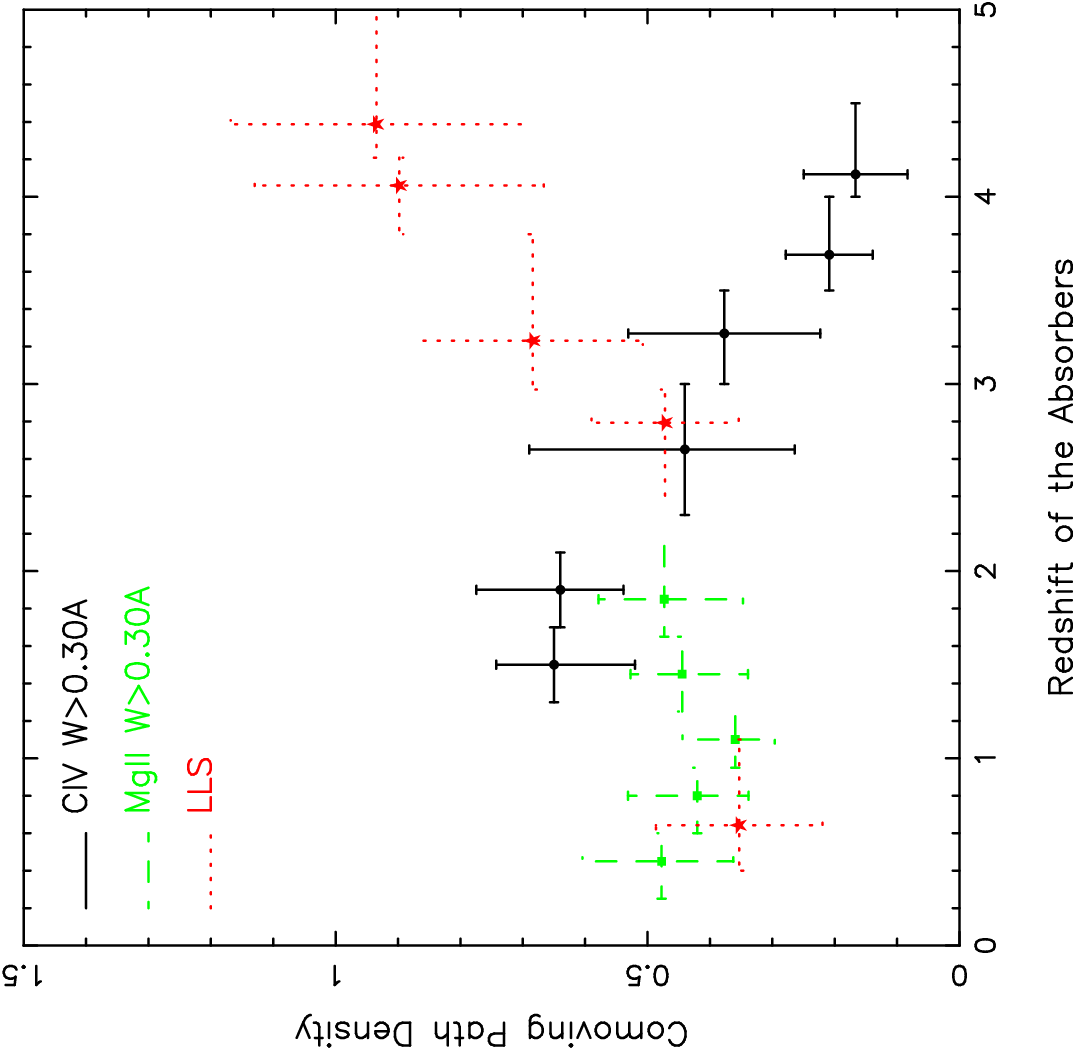}\hfill
    \caption{{\bf Top panel:} number density of various class of
quasar absorbers: $W_{\rm rest}>0.30$\AA\ \civ\ doublets (Steidel
1990, Misawa et al. 2002 and this work according to the redshift
considered), $W_{\rm rest}>0.15$\AA\ \civ\ absorbers (Misawa et
al. 2002), \mgii\ doublets (Steidel \& Sargent 1992) and Lyman-limit
systems (P\'eroux et al. 2003). {\bf Bottom panel:} the comoving path
density: this is the number of systems per comoving path length versus
redshift for our assumed cosmology. A non-evolving population of
absorbers should have a constant comoving path density over time.  }
    \label{f:path}
  \end{center}
\end{figure}

In the bottom panel of Figure~\ref{f:nz}, we present the result of a
similar computation for the $W_{\rm rest}>0.30$\AA\ \mgii\ absorbers
and compare it with the survey of Steidel \& Sargent (1992). Both
studies agree well within the error bars. In Figure~\ref{f:path} (top
panel), we compare the number density of various classes of quasar
absorbers: $W_{\rm rest}>0.30$\AA\ \civ\ doublets (Steidel 1990,
Misawa et al. 2002 and this work according to the redshift
considered), $W_{\rm rest}>0.15$\AA\ \civ\ absorbers (Misawa et
al. 2002), \mgii\ doublets (Steidel \& Sargent 1992) and LLS (P\'eroux
et al. 2003). The first striking point from this figure is the good
agreement between $W_{\rm rest}>0.30$\AA\ \mgii\ absorbers and LLS
number densities in the redshift range where they overlap. Indeed,
both these systems are known to arise from similar regions of galaxies
(Bergeron \& Boiss\'e 1991). Their number density is found to decrease
with decreasing redshift. On the other hand, \civ\ n(z) increases with
time whatever the equivalent width threshold considered. Although, the
number statistics are still small at $W_{\rm rest}>0.15$\AA, it is
clear from Figure~\ref{f:path} that there are at least three times
more of these systems in the Universe than $W_{\rm rest}>0.30$\AA\
absorbers.

The evolution is illustrated in the bottom panel of
Figure~\ref{f:path} where we plot the comoving path density. This is
the number of systems per comoving path length versus redshift for our
assumed cosmology:

\begin{equation}
{\rm path} \hspace{.1cm} {\rm length} = \nonumber \\
                                  n(z) \times (0.3 \times 
				  (1+z)^3+ 0.7 \times (1+z)^2)^{1/2} \times (1+z)^{-2}
\end{equation}

Therefore a non-evolving population of absorbers should have a
constant comoving path density over time. We find that although this
roughly holds for all absorbers at $z<3$, above $z>3.5$ both \civ\
doublets and LLS are found to depart from the horizontal line. In the
region $z=2-3.5$, the error bars are still large for firm conclusions
to be drawn. These results arise from both the new high redshift data
now available and the use of an updated non-zero
$\Lambda$-cosmology. Such an evolution is most probably the signature
of the formation of halos of galaxies of size $\sim$100 h$^{-1}$ kpc
from $\sim$50 h$^{-1}$ kpc systems, a process which seems to end below
$z<3$.

\subsection{Metal Ratios}

\begin{figure}[t]
  \begin{center}
    \includegraphics[width=.25\vsize,height=.9\hsize,angle=-90]{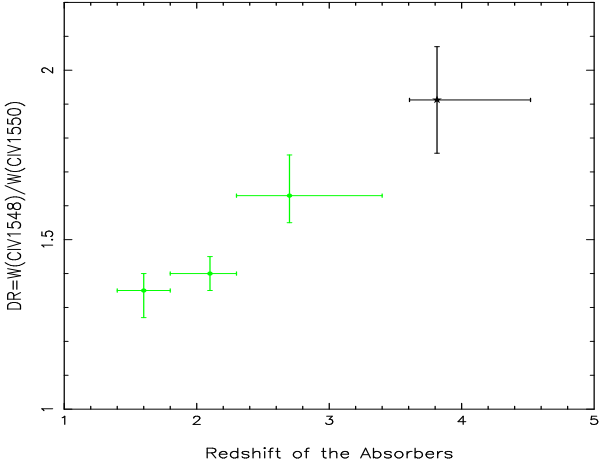}\hfill
    \caption{Redshift evolution of the doublet ratio
    DR=$W(\civ1548)$/$W(\civ1550)$ at $W_{\rm rest}(\civ1548)>0.30$
    \AA. The high-redshift bin is the result from the present work (in
    black). The three low redshift data points (light-coloured) are
    from Steidel (1990).}  \label{f:DR} \end{center}
\end{figure}

The \civ\ doublet ratio DR=$W(\civ1548)$/$W(\civ1550)$ probes the
physical nature of the \civ\ absorbers. Theoretically, DR varies from
2 for the lines situated on the linear part of the curve of growth to
1 for saturated lines. Figure~\ref{f:DR} shows the redshift evolution
of DR by comparing results from Steidel (1990) at low redshift with
results from our sample for \civ\ doublets at $z>3.5$, both with
$W_{\rm rest}(\civ1548)>0.30$ \AA. In order to enable comparison with
previous work, doublets which are separated by less than 1000 km
s$^{-1}$ are merged together and their equivalent width is
combined. This is based on the possibility that clustered components
are not physically independent and that the number of components seen
is sensitive to the spectrum quality. Our resulting sample is composed
of 37 \civ\ doublets with $W_{\rm rest}(\civ1548)>0.30$ \AA\ (note
that this is more than in Table~\ref{t:nber} since in that case it is
$W_{\rm rest}(\civ1550)>0.30$ \AA\ which is used as a criterion). We
note that the trend revealed by Steidel (1990) (i.e. a decrease of DR
with decreasing redshift) already started at earlier times
($z\sim4$). Assuming that DR and W are uncorrelated, this could be due
to either a decrease in the mean Doppler parameter, $b$ or to an
increase in the mean column density, N(\civ), as time goes on. As
pointed out for previous samples (Steidel 1990), we find that DR in
our survey is not correlated with $W_{\rm rest}(\civ1548)$, and
therefore is not expected to be correlated with the line kinematics,
parameterised by $b$ in the range of column densities studied
here. This means that the evolution we probe in Figure~\ref{f:DR} is
probably not due to a change in b, but rather to a systematic increase
in the mean column density with time.

The evolution in the mean column density of \civ\ absorbers could be
interpreted as due to a change in the chemical abundance of carbon
(Songaila 2001 and Pettini et al. 2003). Another viable explanation
for the observed DR evolution is a possible change in ionization state
of the carbon absorbers over time, possibly due to the evolution in
the ionization continuum.

\begin{figure}[t]
  \begin{center}
    \includegraphics[width=.25\vsize,height=.9\hsize,angle=-90]{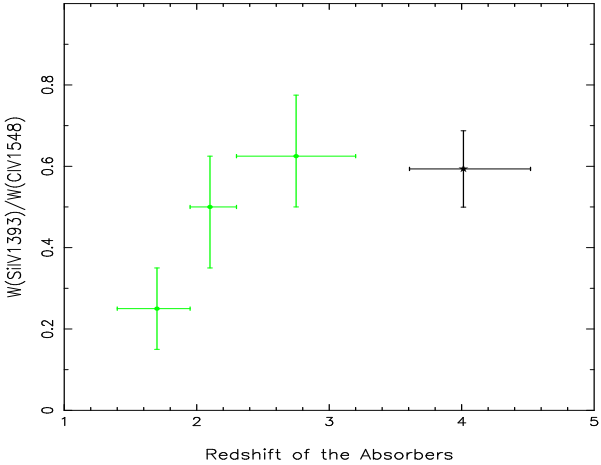}\hfill
    \caption{Redshift evolution of the ratio
    $W$(\siiv1393)/$W$(\civ1548) at $W_{\rm rest}(\civ1548) \ge 0.50$
    \AA. The high-redshift bin is the result from the present work (in
    black). The three low redshift data points (light-coloured) are
    from Bergeron \& Ikeuchi (1990). } \label{f:nber} \end{center}
\end{figure}

In Figure~\ref{f:nber} we plot the ratio $W$(\siiv1393)/$W$(\civ1548)
for the few systems in our sample which contain both these
doublets. This result is shown together with the same measurement
undertaken at lower redshift by Bergeron \& Ikeuchi (1990).  To
facilitate comparison, we limit our sample to pairs with $W_{\rm
rest}(\civ1548) \ge 0.50$ \AA\ which corresponds to only 9 systems. We
find that the ratio $W$(\siiv1393)/$W$(\civ1548) is approximately
constant from $z=3.5$ to $z=2.5$. At $z<3.5$, the trend might be
interpreted as evidence for a systematic increase in the ionization
state of the gas. In other words, this is the opposite trend to what
one would expect given the observed evolution of the \civ\ mean column
density. Nevertheless, it should be emphasized that in order to
compare with previous work from Bergeron \& Ikeuchi (1990), we had to
limit the sample studied to $W_{\rm rest}(\civ1548) \ge 0.50$
\AA. These systems might not be affected by the ionization continuum
in the same way as $W_{\rm rest}(\civ1548) \ge 0.30$ \AA\ lines.
Indeed, other independent lines of evidence based on: absorption
temperature (Schaye et al. 2000); mean continuum depression in quasars
(Bernardi et al. 2003, Theuns et al. 2002); and ionization level of
sensitive elements in quasar absorbers (Vladilo et al. 2003), seem to
favour the scenario where the He II reionization takes place around
$z\sim 3.2$. Therefore, a change of $W$(\siiv1393)/$W$(\civ1548) ratio
is expected around this redshift, either side of which, the ratio
should remain fairly flat, not what we observe in
Figure~\ref{f:nber}. Songaila (1998) found such trends in
high-resolution quasar spectra, but her results were not confirmed by
Kim, Cristiani \& D'Odorico (2002) from comparably high quality data.

To summarise, we observe a decrease of \civ\ doublet DR as time goes
on. This evolution is interpreted as the signature of an increase in
the mean \civ\ column density possibly linked with chemical evolution
or/and with changes in the ionization state of the absorbers.

\section{Conclusion}

We have presented a new sample of high-redshift quasar absorbers
gathered from observations undertaken at the WHT for the northern objects
and at ESO/NTT for the southern ones. These data are analysed in
conjunction with previous studies in order to clarify the statistical
properties of a number of metal lines and link these with the
evolution of galactic structures. Our main findings can be summarised
as follow:

\begin{itemize}

\item{The apparent disagreement between the $n(z)$ redshift evolution
of \civ\ at $W_{\rm rest}(\civ)>0.15$\AA\ reported by Misawa et
al. (2002) from a comparison with Steidel (1990), is the result
of different completeness levels in the absorber samples.}

\item{The number density of $W_{\rm rest}(\civ)>0.30$\AA\ \civ\
doublets increases with decreasing redshift from $z\sim4.5$. $W_{\rm
rest}(\civ)>0.15$\AA\ \civ\ systems are more numerous than the former
at all redshifts. On the other hand, $n(z)$ of $W_{\rm
rest}(\mgii)>0.30$\AA\ \mgii\ doublets and LLS follow each other and
decreases as time goes on.}

\item{At $z>3$, the LLS population is predominant while the number of
$W_{\rm rest}(\civ)>0.30$\AA\ \civ\ is increasing with time. Below
$z<3$, none of all the classes of absorbers show signs of
evolution. We interpret this as the formation of galactic envelopes
from smaller halos.}

\item{The doublet ratio DR=$W(\civ1548)$/$W(\civ1550)$ is found to
decrease with decreasing redshift, a signature of the increase of the
mean \civ\ column density. } 

\item{A study of the $W$(\siiv1393)/$W$(\civ1548) ratio for strong
absorbers ($W_{\rm rest}(\civ1548) \ge 0.50$ \AA) shows a flattening
at $z>3.5$, followed by a decrease, counter to the scenario in which
the DR trend can be explained by a change in ionization continuum.  }

\end{itemize}

\section*{Acknowledgments}
We would like to thank Jacqueline Bergeron for comments on an earlier
version of the manuscript. CP is supported by a Marie Curie
fellowship. This work was supported in part by the European
Communities RTN network "The Physics of the Intergalactic
Medium". This paper is based on observations obtained at the William
Herschel Telescope which is operated on the island of La Palma by the
Isaac Newton Group in the Spanish Observatorio del Roque de los
Muchachos of the Instituto de Astrofisica de Canarias, and on
observations collected during programme ESO 68.A-0279 and ESO
69.A-0541 at the European Southern Observatory with EMMI on NTT
operated at La Silla Observatory, Chile.


\newpage
\begin{onecolumn}

\appendix

\section{Metal Line Identification}

All detected absorptions with confirmed identifications are listed in
Table~\ref{t:metal} below for each quasar spectrum.

\begin{table*}[h]
\begin{center}
\begin{tabular}{lcccccc}
\hline \hline 
Quasars &$z_{\rm em}$ &$\lambda_{\rm obs}$&$z_{\rm abs}$&$W_{\rm obs}$&$W_{\rm rest}$&Identification\\ 
\hline 
BR J0006-6208      & 4.455 &7021.28         &1.956  &2.08  &0.70   &FeII 2374		   \\
                   &       &7046.45         &1.956  &3.55  &1.20   &FeII 2382		   \\
                   &       &7103.95         &3.588  &2.44  &0.53   &CIV 1548			   \\
                   &       &7120.51         &3.588  &1.20  &0.26   &CIV 1550        		   \\
                   &       &7158.94         &3.588  &1.07  &0.23   &CI 1560			   \\
                   &       &7438.79         &1.956  &0.67  &0.23   &SiI 2515			   \\
                   &       &7513.67         &1.687  &0.67  &0.25   &MgII 2796		           \\
                   &       &7533.75         &1.687  &0.16  &0.06   &MgII 2803            	   \\
                   &       &7648.54         &1.956  &3.05  &1.03   &FeII 2586		           \\
                   &       &7686.24         &1.956  &3.96  &1.34   &FeII 2600		           \\
                   &       &8040.22         &1.956  &4.19  &1.42   &if real FeI 2719		   \\
                   &       &8268.55         &1.956  &5.79  &1.96   &MgII 2796 sat		   \\
                   &       &8289.04         &1.956  &5.12  &1.73   &MgII 2803 sat		   \\
                   &       &8435.04         &1.956  &1.22  &0.41   &MgI 2852			   \\
\hline											     
BR J0030$-$5129    & 4.174 &6913.48   &3.465   &1.38  &0.31 &CIV 1548				   \\
                   &       &6925.59   &3.465   &0.96  &0.22 &CIV 1550				   \\
\hline											     
PSS J0034$+$1639   & 4.293 &6663.26         &1.798  &4.71   &1.68    &FeII 2382      		   \\
                   &       &6816.75         &4.225  &0.88   &0.17    &SiII 1304               	   \\
                   &       &6836.35         &3.478  &0.99   &0.22    &SiII 1526       		   \\
                   &       &6877.33         &3.442  &3.07   &0.69    &CIV 1548        		   \\
                   &       &6888.96         &3.442  &2.82   &0.63    &CIV 1550			   \\
                   &       &6935.49         &3.478  &2.74   &0.61    &CIV 1548 (3.442 CI 1560)	   \\
                   &       &6944.17         &3.478  &1.61   &0.36    &CIV 1550 (4.225 CI 1328)	   \\
                   &       &7237.51         &1.588  &2.17   &0.84    &MgII 2796 (1.798 FeII 2586) \\
                   &       &7254.15         &1.588  &0.93   &0.36    &MgII 2803		   \\
                   &       &7279.15         &1.798  &4.61   &1.65    &FeII 2600       		   \\
                   &       &7824.35         &1.798   &4.69  &1.68     &MgII 2796 sat      	   \\
                   &       &7844.70         &1.798   &3.94  &1.41     &MgII 2803		   \\
                   &       &8062.75         &1.883   &2.57  &0.89     &MgII 2796		   \\
                   &       &8082.71         &1.883   &1.04  &0.36     &MgII 2803		   \\
                   &       &8087.25         &4.225   &2.37  &0.45     &CIV 1548			   \\
                   &       &8103.07         &4.225   &1.30  &0.25     &CIV 1550			   \\
\hline											     
PSS J0133$+$0400   & 4.154 &6372.36   &3.771  &5.23  &1.10   &CII 1334        			   \\
                   &       &6406.40   &3.138  &4.57  &1.10   &CIV 1548				   \\
                   &       &6416.68   &3.138  &2.79  &0.67   &CIV 1550				   \\
                   &       &6438.17   &3.620  &1.13  &0.24   &SiIV 1393			   \\
                   &       &6479.49   &3.620  &1.11  &0.24   &SiIV 1402			   \\
                   &       &6507.30   &3.996  &1.63  &0.33   &OI 1302				   \\
                   &       &6514.00   &3.996  &0.53  &0.11   &SiII 1304			   \\
                   &       &6545.49   &3.229  &0.76  &0.18   &CIV 1548 (4.110 CI 1280)		   \\
                   &       &6555.91   &3.229  &0.44  &0.10   &CIV 1550				   \\
                   &       &6648.55   &3.771  &2.11  &0.44   &SiIV 1393			   \\
                   &       &6663.09   &4.110  &1.69  &0.33   &SiII 1304			   \\
                   &       &6691.32   &3.771  &1.11  &0.23   &SiIV 1402  			   \\
                   &       &6894.13   &1.664  &2.36  &0.89   &FeII 2586			   \\
                   &       &6929.30   &1.664  &3.26  &1.22   &FeII 2600       			   \\
                   &       &6962.01   &3.992  &2.88  &0.58   &SiIV 1393			   \\
                   &       &7005.36   &3.992  &1.19  &0.24   &SiIV 1402     			   \\
                   &       &7121.44   &4.110  &0.61  &0.12   &SiIV 1393       			   \\
                   &       &7128.77   &4.115  &1.94  &0.38   &SiIV 1393       			   \\
                   &       &7167.68   &4.110  &0.62  &0.12   &SiIV 1402			   \\
                   &       &7174.99   &4.115  &1.24  &0.24   &SiIV 1402			   \\
                   &       &7281.79   &3.771  &4.46  &0.93   &SiII 1526			   \\
                   &       &7385.12   &3.771  &2.88  &0.60   &CIV 1548				   \\
                   &       &7397.15   &3.771  &1.58  &0.33   &CIV 1550				   \\
\hline\hline
\end{tabular}
\end{center}
\end{table*}

\begin{table*}
\begin{center}
\begin{tabular}{lcccccc}
\hline \hline 
Quasars &$z_{\rm em}$ &$\lambda_{\rm obs}$&$z_{\rm abs}$&$W_{\rm obs}$ &$W_{\rm rest}$&Identification\\ 
\hline 
                   &       &7449.89   &1.664  &4.32  &1.62   &MgII 2796 sat			   \\
                   &       &7469.08   &1.664  &4.11  &1.54   &MgII 2803 sat			   \\
                   &       &7604.01   &1.664  &2.05  &0.77   &MgI 2852       			   \\
                   &       &7671.08   &3.771  &3.10  &0.65   &FeII 1608			   \\
                   &       &7728.49   &3.992  &1.69  &0.34   &CIV 1548				   \\
                   &       &7735.34   &3.996  &1.43  &0.29   &CIV 1548				   \\
                   &       &7742.27   &3.992  &0.89  &0.18   &CIV 1550				   \\
                   &       &7747.94   &3.996  &0.65  &0.13   &CIV 1550				   \\
                   &       &7810.29   &4.115  &0.43  &0.08   &SiII 1526			   \\
\hline					    						     
PSS J0248$+$1802   & 4.422 &6909.82   &1.470  &3.46 &1.40    &MgII 2796       			   \\
                   &       &6926.86   &3.969  &0.95 &0.19    &SiIV 1393 (1.470 MgII 2803)	   \\
                   &       &6970.76   &3.969  &0.30 &0.06    &SiIV 1402			   \\
                   &       &7227.49   &4.185 &1.29  &0.25   &SiIV 1393			   \\
                   &       &7275.61   &4.185 &0.86  &0.17   &SiIV 1402			   \\
\hline				       	               					     
BR J0311$-$1722    & 4.039 &6553.53  &4.034  &0.46  &0.09   &OI 1302				   \\
                   &       &7794.11  &4.034  &6.63  &1.32   &CIV 1548 assym profiles 		   \\
                   &       &7808.10  &4.034  &2.37  &0.47   &CIV 1550				   \\
\hline				       	               					     
BR J0324$-$2918    & 4.622 &7347.10   &3.746  &0.57  &0.12   &CIV 1548				   \\
                   &       &7359.84   &3.746  &0.72  &0.15   &CIV 1550        			   \\
                   &       &7564.20   &2.227  &0.91  &0.28   &FeII 2344			   \\
                   &       &7691.00   &2.227  &1.42  &0.44   &FeII 2382       			   \\
                   &       &8176.73   &1.923  &1.33  &0.46   &MgII 2796			   \\
                   &       &8196.15   &1.923  &0.64  &0.22   &MgII 2803			   \\
                   &       &8345.80   &2.227  &0.77  &0.24   &FeII 2586			   \\
                   &       &8392.62   &2.227  &1.36  &0.42   &FeII 2600			   \\
\hline				       	                      					     
BR J0334$-$1612    & 4.363 &6673.26   &3.788  &1.05  &0.22   &SiIV 1393			   \\
                   &       &6713.80   &3.788  &0.57  &0.12   &SiIV 1402			   \\
                   &       &7228.38   &3.668  &1.07  &0.23   &CIV 1548 (4.186 SiIV 1393)	   \\
                   &       &7239.12   &3.668  &0.37  &0.08   &CIV 1550 			   \\
                   &       &7275.85   &4.186  &1.93  &0.37   &SiIV 1402			   \\
                   &       &7368.97   &3.759  &1.61  &0.34   &CIV 1548				   \\
                   &       &7381.79   &3.759  &0.81  &0.17   &CIV 1550        			   \\
\hline				       	                      					     
BR J0355$-$3811    & 4.545 &7516.08   &3.855  &0.51  &0.11   &CIV 1548				   \\
                   &       &7528.44   &3.855  &0.59  &0.12   &CIV 1550				   \\
                   &       &7723.51   &1.986  &2.09  &0.70   &FeII 2586 			   \\
                   &       &7764.48   &1.986  &3.49  &1.17   &FeII 2600       			   \\
                   &       &8226.45   &4.313  &2.88  &0.54   &CIV 1548				   \\
                   &       &8233.01   &4.318  &3.92  &0.74   &CIV 1548				   \\
                   &       &8239.86   &4.313  &1.02  &0.19   &CIV 1550				   \\
                   &       &8248.49   &4.318  &1.06  &0.20   &CIV 1550				   \\
                   &       &8349.70   &1.986  &10.28 &3.42   &MgII 2796			   \\
                   &       &8371.16   &1.986  &9.38  &3.14   &MgII 2803       			   \\
\hline				       	                      					     
BR J0419$-$5716    & 4.461 &7217.62   &2.028  &2.45  &0.81   &FeII 2382               		   \\
                   &       &7834.70   &2.028  &1.51  &0.50   &FeII 2586			   \\
                   &       &7875.67   &2.028  &2.30  &0.76   &FeII 2600       			   \\
                   &       &8080.50   &4.218  &2.73  &0.52   &CIV 1548				   \\
                   &       &8095.89   &4.218  &1.49  &0.29   &CIV 1550        			   \\
                   &       &8225.50   &4.311  &3.32  &0.63   &CIV 1548				   \\
                   &       &8239.86   &4.311  &0.97  &0.18   &CIV 1550        			   \\
                   &       &8411.29   &4.434  &0.60  &0.11   &CIV 1548				   \\
                   &       &8426.32   &4.434  &0.44  &0.08   &CIV 1550        			   \\
                   &       &7132.42   &3.606  &5.20  &1.13   &CIV 1548				   \\
                   &       &7143.72   &3.606  &3.25  &0.71   &CIV 1550				   \\
                   &       &7209.92   &4.172  &2.08  &0.40   &SiIV 1393 			   \\
                   &       &7249.67   &3.682  &1.07  &0.23   &CIV 1548  			   \\
                   &       &7255.61   &4.172  &2.10  &0.41   &SiIV 1402 			   \\
                   &       &7261.58   &3.682  &0.82  &0.18   &CIV 1550  			   \\
                   &       &8007.87   &4.172  &3.14  &0.61   &CIV 1548				   \\
                   &       &8020.81   &4.172  &1.88  &0.36   &CIV 1550				   \\
\hline
BR J0426$-$2202    & 4.320 &7132.42   &3.606  &5.20  &1.13   &CIV 1548   \\ 
                   &       &7143.72   &3.606  &3.25  &0.71   &CIV 1550   \\
\hline\hline
\end{tabular}
\end{center}
\end{table*}

\begin{table*}[!h]
\begin{center}
\begin{tabular}{lcccccc}
\hline \hline 
Quasars &$z_{\rm em}$ &$\lambda_{\rm obs}$&$z_{\rm abs}$&$W_{\rm obs}$&$W_{\rm rest}$&Identification\\ 
\hline 
                   &       &7209.92   &4.172  &2.08  &0.40   &SiIV 1393  \\ 
                   &       &7249.67   &3.682  &1.07  &0.23   &CIV 1548   \\
                   &       &7255.61   &4.172  &2.10  &0.41   &SiIV 1402  \\
                   &       &7249.67   &3.682  &1.07  &0.17   &CIV 1548   \\
                   &       &8007.87   &4.172  &3.14  &0.61   &CIV 1548   \\ 
                   &       &8020.81   &4.172  &1.88  &0.36   &CIV 1550   \\ 
\hline
PMN J0525$-$3343   & 4.383 &7057.43   &4.063  &0.65  &0.13   &SiIV 1393			   \\
                   &       &7093.42   &3.581  &1.00  &0.22   &CIV 1548 (4.063 SiIV 1402)	   \\
                   &       &7104.99   &3.581  &0.78  &0.17   &CIV 1550        			   \\
                   &       &7182.54   &1.568  &3.05  &1.19   &MgII 2796			   \\
                   &       &7201.04   &1.568  &3.19  &1.24   &MgII 2803 			   \\
                   &       &7367.39   &3.581  &0.92  &0.20   &FeII 1608			   \\
                   &       &7569.07   &4.430  &3.40  &0.63   &SiIV 1393			   \\
                   &       &7618.63   &4.430  &4.57  &0.84   &SiIV 1402			   \\
                   &       &7729.29   &4.063  &1.36  &0.27   &SiII 1526       			   \\
                   &       &7841.89   &4.063  &1.45  &0.29   &CIV 1548				   \\
                   &       &7854.16   &4.063  &0.68  &0.13   &CIV 1550				   \\
                   &       &8145.68   &4.063  &0.80  &0.16   &FeII 1608			   \\
                   &       &225.917   &4.313  &3.48  &0.65   &CIV 1548				   \\
                   &       &8239.63   &4.313  &0.46  &0.09   &CIV 1550				   \\
\hline				       	                      					     
BR J0529$-$3552    & 4.172 &6605.73   &4.062  &1.42  &0.28   &SiII 1304       			   \\
                   &       &6708.94   &1.398  &1.20  &0.50   &MgII 2796			   \\
                   &       &6726.52   &1.398  &0.57  &0.24   &MgII 2803 			   \\
                   &       &6759.82   &4.062  &0.49  &0.10   &CII 1334				   \\
                   &       &6779.25   &1.423  &0.87  &0.36   &MgII 2796			   \\
                   &       &6795.07   &1.423  &0.77  &0.32   &MgII 2803                	   \\
                   &       &6973.94   &3.503  &2.40  &0.53   &CIV 1548				   \\
                   &       &6985.50   &3.503  &1.39  &0.31   &CIV 1550				   \\
                   &       &7058.81   &4.062  &1.84  &0.36   &SiIV 1393 			   \\
                   &       &7103.20   &4.062  &1.45  &0.29   &SiIV 1402			   \\
                   &       &7244.21   &3.503  &1.48  &0.33   &FeII 1608               		   \\
                   &       &7427.65   &1.656  &0.91  &0.34   &MgII 2796			   \\
                   &       &7449.85   &1.656  &1.20  &0.45   &MgII 2803			   \\
                   &       &7528.19   &3.503  &1.00  &0.22   &AlII 1670       			   \\
                   &       &7729.91   &4.062  &0.94  &0.19   &SiII 1526			   \\
\hline				       	                       					     
BR J0714$-$6455    & 4.462 &7238.83   &4.193  &0.85  &0.16   &SiIV 1393    			   \\
                   &       &7248.78   &3.747  &1.03  &0.22   &SiII 1526 (4.193 SiIV 1402)         \\
                   &       &7350.31   &3.747  &1.97  &0.41   &CIV 1548				   \\
                   &       &7362.34   &3.747  &0.74  &0.16   &CIV 1550				   \\
                   &       &7689.35   &3.966  &2.58  &0.52   &CIV 1548				   \\
                   &       &7701.69   &3.966  &2.04  &0.41   &CIV 1550				   \\
                   &       &8040.53   &4.193  &2.55  &0.49   &CIV 1548				   \\
                   &       &8053.67   &4.193  &1.17  &0.23   &CIV 1550				   \\
\hline				       	                      					     
PSS J0747$+$4434   & 4.430 &...          &...       &...      &...       &...					   \\
\hline				       	                      					     
PSS J1159$+$1337   & 4.073 &6418.61   &1.739  &1.09  &0.40   &FeII 2344       			   \\
                   &       &6502.11   &1.739  &0.43  &0.16   &FeII 2374			   \\
                   &       &6522.56   &1.739  &1.83  &0.67   &FeII 2382			   \\
                   &       &6582.48  &3.724   &1.24  &0.26   &SiIV 1393               		   \\
                   &       &6625.76  &3.756   &1.11  &0.23   &SiIV 1393 (3.724 SiIV 1402)	   \\
                   &       &6670.04  &3.756   &0.34  &0.07   &SiIV 1402			   \\
                   &       &7081.73  &1.739   &0.75  &0.27   &FeII 2586     			   \\
                   &       &7119.62  &1.739   &1.64  &0.60   &FeII 2600			   \\
                   &       &7311.64  &3.724   &1.02  &0.22   &CIV 1548				   \\
                   &       &7323.59  &3.724   &0.42  &0.09   &CIV 1550				   \\
\hline				      	                       					     
PSS J1253$+$0228   & 4.007 &6142.86  &3.606   &1.23  &0.27   &CII 1334				   \\
                   &       &6320.60  &1.261   &2.22  &0.98   &MgII 2796			   \\
                   &       &6335.10  &1.261   &0.89  &0.39   &MgII 2803       			   \\
                   &       &6415.92  &3.606   &4.98  &1.08   &SiIV 1393			   \\
                   &       &6458.09  &3.606   &3.89  &0.84   &SiIV 1402			   \\
                   &       &7127.23  &3.606   &5.25  &1.14   &CIV 1548        			   \\
                   &       &7139.24  &3.606   &4.16  &0.90   &CIV 1550				   \\
\hline\hline
\end{tabular}
\end{center}
\end{table*}

\begin{table*}[!h]
\begin{center}
\begin{tabular}{lcccccc}
\hline \hline 
Quasars &$z_{\rm em}$ &$\lambda_{\rm obs}$&$z_{\rm abs}$&$W_{\rm obs}$&$W_{\rm rest}$&Identification\\ 
\hline				      	                       					     
BR J1330$-$2522    & 3.949 &6319.29  &3.084   &2.01  &0.49   &CIV 1548				   \\
                   &       &6329.36  &3.084   &1.51  &0.37   &CIV 1550				   \\
                   &       &6645.79  &3.772   &0.78  &0.16   &SiIV 1393			   \\
                   &       &6689.87  &3.772   &0.32  &0.07   &SiIV 1402			   \\
                   &       &6794.58  &3.390   &0.85  &0.19   &CIV 1548				   \\
                   &       &6805.62  &3.390   &0.81  &0.18   &CIV 1550				   \\
                   &       &6819.10  &3.084   &1.38  &0.34   &AlII 1670			   \\
                   &       &6874.38  &3.440   &1.55  &0.35   &triplet				   \\
                   &       &7383.72  &3.769   &1.93  &0.40   &CIV 1548				   \\
                   &       &7395.73  &3.769   &1.64  &0.34   &CIV 1550				   \\
\hline 
PSS 1456$+$2007    & 4.249 &7720.09  &1.761   &0.95  &0.34   &MgII 2796 			   \\
                   &       &7740.89  &1.761   &0.70  &0.25   &MgII 2803			   \\
                   &       &7796.79  &4.036   &1.25  &0.25   &CIV 1548				   \\
                   &       &7807.97  &4.036   &0.58  &0.12   &CIV 1550				   \\
\hline 						      0.11   
PSS J1618$+$4125   & 4.213 &7276.55  &4.223   &0.60  &       &SiIV 1393			   \\
                   &       &7382.46  &4.223   &0.57  &0.11   &SiIV 1402			   \\
\hline				      	              0.16     					     
PSS J1633$+$1411   & 4.351 &7091.98  &3.580   &0.75  &       &CIV 1548				   \\
                   &       &7104.05  &3.580   &0.40  &0.09   &CIV 1550				   \\
                   &       &7179.44  &4.150   &0.96  &0.19   &SiIV 1393			   \\
                   &       &7225.18  &4.150   &0.70  &0.14   &SiIV 1402			   \\
                   &       &8177.34  &4.282   &2.37  &0.45   &CIV 1548 			   \\
                   &       &8192.78  &4.282   &1.573 &0.30   &CIV 1550 			   \\
\hline				   		             					     
PSS J1646$+$5514   & 4.037 &6988.64  &3.516   &0.23  &0.05   &CIV 1548				   \\
                   &       &7000.70  &3.516   &0.33  &0.07   &CIV 1550 (3.521 CIV 1548)	   \\
                   &       &7010.49  &4.030   &0.89  &0.18   &SiIV 1393 (3.521 CIV 1550)	   \\
                   &       &7033.53  &3.544   &2.26  &0.50   &CIV 1548        			   \\
                   &       &7045.50  &3.544   &1.40  &0.31   &CIV 1550 			   \\
                   &       &7055.81  &4.030   &0.48  &0.10   &SiIV 1402 			   \\
                   &       &7357.65  &3.754   &1.43  &0.30   &CIV 1548 			   \\
                   &       &7366.80  &3.759   &0.40  &0.08   &CIV 1548				   \\
                   &       &7369.19  &3.754   &1.11  &0.23   &CIV 1550				   \\
                   &       &7378.21  &3.759   &0.14  &0.03   &CIV 1550				   \\
                   &       &7678.92  &4.030   &0.65  &0.13   &SiII 1526			   \\
                   &       &7786.91  &4.030   &1.39  &0.28   &CIV 1548                  	   \\
                   &       &7797.52  &4.037   &1.28  &0.25   &CIV 1548				   \\
                   &       &7798.54  &4.030   &2.38  &0.47   &CIV 1550  			   \\
                   &       &7809.73  &4.037   &0.44  &0.09   &CIV 1550                  	   \\
                   &       &7994.39  &1.859   &0.92  &0.32   &MgII 2796			   \\
                   &       &8015.44  &1.859   &0.51  &0.18   &MgII 2803			   \\
                   &       &8089.58  &4.030   &0.14  &0.03   &FeII 1608			   \\
\hline				   		             					     
PSS J1723$+$2243   & 4.520 &7065.98  &3.696   &0.69  &0.15   &SiII 1526 (dbl cpt)		   \\
                   &       &7283.89  &3.704   &3.02  &0.64   &CIV 1548				   \\
                   &       &7296.29  &3.704   &1.29  &0.27   &CIV 1550        			   \\
                   &       &7311.00  &4.244   &1.20  &0.23   &SiIV 1393			   \\
                   &       &7355.29  &4.244   &0.49  &0.09   &SiIV 1402			   \\
                   &       &7550.54  &3.696   &1.51  &0.32   &FeII 1608 (dbl cpt)		   \\
                   &       &7843.36  &3.696   &2.67  &0.57   &AlII 1670 (dbl cpt)		   \\
                   &       &8004.04  &4.244   &0.83  &0.16   &SiII 1526			   \\
                   &       &8104.95  &4.234   &0.85  &0.16   &CIV 1548				   \\
                   &       &8117.52  &4.244   &1.81  &0.35   &CIV 1548				   \\
                   &       &8120.59  &4.234   &1.52  &0.29   &CIV 1550 			   \\
                   &       &8137.92  &4.244   &0.77  &0.15   &CIV 1550				   \\
\hline				      	                      					     
PSS J1802$+$5616   & 2.891 &7035.30  &4.048   &0.69  &0.14   &SiIV 1393			   \\
                   &       &7080.97  &4.048   &0.50  &0.10   &SiIV 1402 (3.651 SiII 1526)	   \\
                   &       &8035.92  &4.190   &1.33  &0.26   &  CIV 1548			   \\
                   &       &8047.76  &4.190   &0.44  &0.08   &  CIV 1550			   \\
\hline\hline
\end{tabular}
\end{center}
\end{table*}

\begin{table*}[!h]
\begin{center}
\begin{tabular}{lcccccc}
\hline \hline 
Quasars &$z_{\rm em}$ &$\lambda_{\rm obs}$&$z_{\rm abs}$&$W_{\rm obs}$&$W_{\rm rest}$&Identification\\ 
\hline				      	                      					     
PSS J2154$+$0335   & 4.363 &6912.21  &3.961   &1.07  &0.22   &SiIV 1393 			   \\
                   &       &6958.06  &3.961   &0.79  &0.16   &SiIV 1402			   \\
                   &       &7127.54  &1.756   &1.72  &0.62   &FeII 2586			   \\
                   &       &7168.48  &1.756   &3.33  &1.21   &FeII 2600			   \\
                   &       &7531.09  &3.776   &0.89  &0.19   &CIV 1548				   \\
                   &       &7545.17  &3.776   &1.00  &0.21   &CIV 1550 cont        		   \\
                   &       &7681.74  &3.961   &1.98  &0.40   &CIV 1548				   \\
                   &       &7694.48  &3.961   &1.47  &0.30   &CIV 1550				   \\
                   &       &7706.41  &1.756   &4.68  &1.70   &MgII 2796 sat			   \\
                   &       &7727.70  &1.756   &4.49  &1.63   &MgII 2803 sat      		   \\
                   &       &7865.25  &1.756   &1.07  &0.39   &MgI 2852				   \\
\hline 				      	              					     
PSS J2155$+$1358   & 4.256 &6943.22  &1.914  &2.23  &0.77   &FeII 2382			   \\
                   &       &7067.67  &3.567  &2.98  &0.65   &CIV 1548				   \\
                   &       &7079.20  &3.567  &2.02  &0.44   &CIV 1550				   \\
                   &       &7308.18  &4.244  &2.15  &0.41   &SiIV 1393       			   \\
                   &       &7354.97  &4.244  &2.28  &0.43   &SiIV 1402			   \\
                   &       &7538.26  &1.914  &0.42  &0.14   &FeII 2586			   \\
                   &       &7579.07  &1.914  &1.33  &0.46   &FeII 2600			   \\
                   &       &8117.54  &4.244  &2.90  &0.55   &  CIV 1548 sat			   \\
                   &       &8131.14  &4.244  &2.22  &0.42   &  CIV 1550 sat			   \\
                   &       &8148.42  &1.914  &5.16  &1.77   &  MgII 2796 sat			   \\
                   &       &8170.12  &1.914  &5.08  &1.74   &  MgII 2803 sat			   \\
\hline				      	                      					     
BR J2216$-$6714    & 4.469 &7173.70  &2.060  &3.29  &1.08   &FeII 2344       			   \\
                   &       &7268.79  &2.060  &2.43  &0.79   &FeII 2374			   \\
                   &       &7295.82  &2.060  &3.61  &1.18   &FeII 2382       			   \\
                   &       &7320.39  &3.728  &0.82  &0.17   &CIV 1548        			   \\
                   &       &7333.77  &3.728  &0.70  &0.15   &CIV 1550				   \\
                   &       &7367.07  &4.285  &0.84  &0.16   &SiIV 1393			   \\
                   &       &7413.73  &4.285  &0.68  &0.13   &SiIV 1402			   \\
                   &       &7491.72  &3.837  &0.64  &0.13   &CIV 1548				   \\
                   &       &7503.76  &3.837  &0.47  &0.10   &CIV 1550				   \\
                   &       &7607.87  &3.728  &0.40  &0.08   &FeII 1608       			   \\
                   &       &7889.88  &4.095  &1.30  &0.26   &CIV 1548				   \\
                   &       &7902.97  &4.095  &0.33  &0.06   &CIV 1550				   \\
                   &       &7914.40  &2.060  &3.09  &1.01   &FeII 2586			   \\
                   &       &7956.90  &2.060  &4.44  &1.45   &FeII 2600			   \\
\hline				      	                      					     
PSS J2344$+$0342   & 4.239 &6531.04  &3.219  &2.20  &0.52   &CIV 1548				   \\
                   &       &6542.17  &3.219  &2.04  &0.48   &CIV 1550				   \\
                   &       &6589.50  &4.053  &0.88  &0.17   &SiII 1304			   \\
                   &       &6789.50  &3.219  &1.81  &0.43   &FeII 1608       			   \\
                   &       &6805.74  &3.882  &1.52  &0.31   &SiIV 1393			   \\
                   &       &6849.34  &3.882  &1.31  &0.27   &SiIV 1402       			   \\
                   &       &7050.98  &3.219  &2.93  &0.69   &AlII 1670       			   \\
                   &       &7453.44  &3.882  &0.75  &0.15   &SiII 1526			   \\
                   &       &7559.01  &3.882  &2.67  &0.55   &CIV 1548			   \\
                   &       &7571.64  &3.882  &1.56  &0.32   &CIV 1550				   \\
                   &       &7714.09  &4.053  &0.54  &0.11   &SiII 1526			   \\
                   &       &7824.45  &4.053  &0.43  &0.09   &CIV 1548				   \\
                   &       &7836.06  &4.053  &0.22  &0.04   &CIV 1550				   \\
\hline				      	                      					     
BR J2349$-$3712    & 4.208 &6538.49  &3.690  &1.23  &0.26   &SiIV 1393 (3.284 SiII 1526)	   \\
                   &       &6581.17  &3.690  &0.93  &0.20   &SiIV 1402			   \\
                   &       &6592.52  &3.258  &2.20  &0.52   &CIV 1548				   \\
                   &       &6603.81  &3.258  &1.34  &0.31   &CIV 1550				   \\
                   &       &6632.13  &3.284  &0.66  &0.15   &CIV 1548				   \\
                   &       &6643.93  &3.284  &0.54  &0.13   &CIV 1550				   \\
                   &       &6891.96  &3.284  &0.99  &0.23   &FeII 1608			   \\
                   &       &7161.63  &3.284  &0.77  &0.18   &AlII 1670       			   \\
\hline\hline
\end{tabular}
\end{center}
\end{table*}

\begin{table*}[!h]
\begin{center}
\begin{tabular}{lcccccc}
\hline \hline 
Quasars &$z_{\rm em}$ &$\lambda_{\rm obs}$&$z_{\rm abs}$&$W_{\rm obs}$&$W_{\rm rest}$&Identification\\ 
\hline
                   &       &7267.37  &3.690  &1.06  &0.23   &CIV 1548				   \\
                   &       &7274.45  &3.690  &0.78  &0.17   &CIV 1550				   \\
                   &       &7364.09  &3.757  &2.13  &0.45   &CIV 1548				   \\
                   &       &7377.16  &3.757  &1.60  &0.34   &CIV 1550				   \\
                   &       &7679.67  &3.960  &1.54  &0.31   &CIV 1548				   \\
                   &       &7694.11  &3.960  &0.61  &0.12   &CIV 1550				   \\
                   &       &7714.71  &1.758  &1.01  &0.37   &MgII 2796       			   \\
                   &       &7734.25  &1.758  &0.35  &0.13   &MgII 2803			   \\
                   &       &7832.95  &3.690  &1.30  &0.28   &AlII 1670			   \\
\hline\hline
\end{tabular}
\end{center}
\scriptsize{$^*$Object possibly affect d by BAL features.}
\caption{List of detected absorption lines which are successfully
identified with observed wavelength, equivalent width, and
corresponding absorption redshift.}
\label{t:metal}
\end{table*}

\end{onecolumn}

\end{document}